\documentclass{aa}
\usepackage[varg]{txfonts}

\title{A distance measurement for blazar TXS~0506+056\\using its radio variability and very long baseline images}
\titlerunning{A distance measurement for blazar TXS~0506+056}

\author{
Chanwoo~Song\inst{1,2}
\and Sang-Sung~Lee\inst{1,2}\thanks{Corresponding author: sslee@kasi.re.kr}
\and Sincheol~Kang\inst{1}
\and Whee~Yeon~Cheong\inst{1,2}
}
\authorrunning{Song~C. et al.}

\institute{
Korea Astronomy and Space Science Institute, 776 Daedeok-daero, Daejeon, 34055, Korea
\and University of Science and Technology, 217 Gajeong-ro, Yuseong-gu, Daejeon, 34113, Korea
}

\date{Received February~10,~2025 / Accepted November~14, 2025}

\begin{document}

\abstract
{}
{We present the results of constraining the angular diameter distance to blazar TXS~0506+056 ($z=0.3365$), a radio-bright active galactic nucleus (AGN) whose jet is aligned with the line of sight.}
{We used data obtained with the 15~GHz Very Long Baseline Array (VLBA) from MJD~54838 to MJD~60262 (15~years) and data from the 15~GHz Owens Valley Radio Observatory (OVRO) 40~m single dish (SD) telescope from MJD~54474 to MJD~59023 (12~years).
We used a variability timescale and a causality argument of a linear size (taking the Doppler factor and a cosmological redshift into account) to measure the angular diameter distance to the source. To constrain the Doppler factor, we applied the relation between the rest-frame brightness temperature of the emission region and the observed brightness temperature.
To calculate the observed brightness temperature, the angular size and flux density variation of the emission region are required.
The angular size of the emission region (i.e., the VLBA core) was obtained from a full width at half maximum, which is a circular Gaussian model-fitting parameter that ranges from 0.048--0.228 mas, and its uncertainty is determined to be 1.8--13~\%.
Using the OVRO~SD light curve, we obtained a {variability timescale of $\tau=128.0_{-0.3}^{+0.2}$~days and a peak flux density of $1.750_{-0.104}^{+0.015}$~Jy for the largest flare that peaked on MJD~$58921.7_{-5.5}^{+2.6}$.}
We assumed a disk brightness geometry, equipartition brightness temperature ($T_{\rm b,int}=5\times10^{10}$~K), and perfect radius.}
{By fitting the circular Gaussian model to the VLBA images, we found that the variability in the VLBA core drives the multiple flares.
Based on the timescales and peak flux densities for the flares, we calculated the angular diameter distance.
Using the VLBA core sizes obtained near the flare peaks, we found consistent distance measurement results with the $\Lambda$CDM model within 1$\sigma$ uncertainties.}
{We suggest that the best distance from the source is $941_{-64}^{+59}$~Mpc, which is comparable with the $\Lambda$CDM distance of $948.2\pm13.5$~Mpc.
The distance measurement should indeed be taken at the peak of a flare.
We found that the decomposed timescale allowed us to obtain consistent distances with the $\Lambda$CDM. We strongly suggest to decompose light curves when the variability timescales are to be obtained properly.}


\maketitle

\makeatletter
\let\linenumbers\relax
\makeatother
\ifdefined\nolinenumbers\nolinenumbers\fi

\section{Introduction}\label{sect:1}
The lambda cold dark matter ($\Lambda$CDM) model is a well-constrained cosmological model for explaining the expansion of the Universe and for estimating cosmological distances.
For the $\Lambda$CDM model without radiation and spatial curvature, the Hubble parameter ($H$) is written as
\begin{equation}
    H(z)=H_{0}\sqrt{\Omega_{\rm m}(1+z)^{3}+\Omega_{\rm \Lambda}},
    \label{eq:Hubble_parameter}
\end{equation}
where $z$ is the redshift, $H_{0}$ is the Hubble constant, $\Omega_{\rm m}$ is the matter density parameter including the baryonic and cold dark matter, and $\Omega_{\rm \Lambda}$ is the dark energy density parameter related to $\Omega_{\Lambda}\sim1-\Omega_{\rm m}$.
The cosmological distance is related to $z$ as
\begin{equation}
    D_{\rm L}=D_{\rm C}(1+z)=D_{\rm A}(1+z)^{2}=c(1+z)\int_{0}^{z}\frac{{\rm d}z'}{H(z')},
    \label{eq:cosmic_distances}
\end{equation}
where $D_{\rm L}$ is the luminosity distance, $D_{\rm C}$ is the comoving distance, and $D_{\rm A}$ is the angular diameter distance~\citep{Hogg1999}.
To infer the cosmological parameters $H_{0}$, $\Omega_{\rm m}$, and $\Omega_{\rm \Lambda}$, the cosmic distance ladder can be used. This is broadly divided into two categories: the direct distance ladder, and the inverse distance ladder.
The direct distance ladder determines the distance from specific celestial objects such as Cepheid variables, Type Ia supernovae (SNe~Ia), and the red giant branch (RGB; i.e., tip of the RGB).
By fitting Eq.~(\ref{eq:cosmic_distances}) to the distance--redshift relation of the direct distance ladder, the cosmological parameters can be inferred.
The inverse distance ladder infers the cosmological parameters by observing early-time processes such as the cosmic microwave background (CMB) and the baryon acoustic oscillation (BAO).
The temperature anisotropy observed in the CMB reflects density fluctuations in the early universe.
The BAO is an acoustic density wave created by fluctuations in the density of baryonic matter in the plasma state of the early universe.
The direct and inverse distance ladders determine the cosmological parameters independently.
 $H_{0}$ as constrained by the two distance ladders differs, however. This is called the Hubble tension.
The direct distance ladder constrains $H_{0}=73.04\pm1.04$~km/s/Mpc~\citep{Riess2022}, and the inverse distance ladder constrains $H_{0}=67.4\pm0.5$~km/s/Mpc~\citep{Aghanim2020}.
Solving the Hubble tension currently is a great challenge.
One of the tasks for solving the Hubble tension is expanding the local limits of the direct distance ladder because the direct distance ladder clearly has a fatal flaw at higher redshifts.
Among the rungs of the direct distance ladder, SNe~Ia are known to be the farthest. They reach $z=2.26$~\citep{Scolnic2018}.
It is difficult to observe distant objects because the intensity dims by $(1+z)^{3}$ as a function of $z$.
The direct distance ladder is then accurate to the current and the local universe.
Finding new rungs of the ladder with a higher redshift might be a clue to verify whether the locality of the direct distance ladder is the cause of the Hubble tension.\\
\indent Blazars are a subclass of active galactic nuclei (AGNs) in which relativistic jets powered by a supermassive black hole (SMBH) at the center are observed, being closely aligned to the line of sight.
When the relativistic particles in the jet pass through the magnetic field, synchrotron radiation is emitted in a wide range of electromagnetic waves from radio to gamma-rays.
Synchrotron radiation is Doppler-boosted by the relativistic effect by $\delta^{3}$ as a function of the Doppler factor $\delta$, and it can therefore be observed from distant blazars at high redshift.
In particular, very long baseline interferometry (VLBI) enables us to obtain high-resolution images of the blazars on parsecond scales and to measure accurate angular sizes.
High-redshift blazars above 6.10 can be imaged by the VLBI~\citep{Zhang2022}.
 \citet{Hodgson2020} applied a method of distance determination using the timescale--size causality to AGN 3C~84, which is not Doppler-boosted ($\delta\sim1$). They obtained a consistent distance measurement with SNe~Ia and Tully--Fisher measurements.
This research has led to the expectation that AGNs might be a direct distance ladder.
This method has not yet been applied to a highly Doppler-boosted ($\delta>1$) source, however. 
 \citet{Hodgson2023} suggested a method of distance determination using Doppler-boosted celestial objects without the Doppler factor.
The blazar TXS~0506+056 is classified as an intermediate spectrally peaked BL~Lac object AGN ($z=0.3365$).
On May~13, 2020, the 15~GHz radio flux density reached a maximum of 2.44~Jy and was observed by the Owens Valley Radio Observatory (OVRO) 40~m radio telescope~\citep{Hovatta2021}.
 $\delta$ rose to a maximum of 13.6 on June~17, 2017 \citep{Li2020}.
We applied the method to TXS~0506+056 to examine whether blazars can be used as a valid means for distance measurements.\\
\indent In Sect.~\ref{sect:2} we derive the angular diameter distance formula of blazars using the variability timescale, flux density, and angular sizes of the emission region.
In Sect.~\ref{sect:3} we introduce the data and analysis methods we used to obtain the parameters for the distance measurements.
In Sect.~\ref{sect:4} we describe the analysis results and distance measurements.
In Sect.~\ref{sect:5},we discuss the results, and in Sect.~\ref{sect:6} we conclude.

\section{Background}\label{sect:2}
In this section, we describe the method presented by~\citet{Hodgson2023} for introducing scaling factors ($K$, $M$, and their combination $KM^{3}$; see below for details), which will also be introduced in~\citet{Cheong2025}.
We used superscripts $Q^{\rm em}$, $Q^{\rm so}$, and $Q^{\rm rec}$ over an arbitrary quantity $Q$, corresponding to the emission, source, and receiver reference frames, respectively. The emission frame is referenced to the relativistic jet of a blazar. The source frame is referenced to the host galaxy of the blazar.
\begin{table}[!t]
    \caption{Physical quantity ($Q$) transformation by Doppler effect and redshift.}
    \label{tab:frame_transform}
    \centering
    \begin{tabular}{@{\extracolsep{0pt}}lll}
        \hline\hline
          $Q^{\rm so}$ to $Q^{\rm em}$
        & $Q^{\rm rec}$ to $Q^{\rm so}$
        & $Q^{\rm rec}$ to $Q^{\rm em}$
        \\
        \hline
          $\nu^{\rm so}=\delta\nu^{\rm em}$
        & $\nu^{\rm rec}=\nu^{\rm so}/(1+z)$
        & $\nu^{\rm rec}=\delta\nu^{\rm em}/(1+z)$ 
        \\
          ${\rm d}t^{\rm so}={\rm d}t^{\rm em}/\delta$ 
        & ${\rm d}t^{\rm rec}=(1+z){\rm d}t^{\rm so}$
        & ${\rm d}t^{\rm rec}=(1+z){\rm d}t^{\rm em}/\delta$
        \\
          ${\rm d}A^{\rm so}={\rm d}A^{\rm em}/\delta^{2}$
        & ${\rm d}A^{\rm rec}={\rm d}A^{\rm so}$
        & ${\rm d}A^{\rm rec}={\rm d}A^{\rm em}/\delta^{2}$
        \\
          ${\rm d}\theta^{\rm so}={\rm d}\theta^{\rm em}$
        & ${\rm d}\theta^{\rm rec}={\rm d}\theta^{\rm so}$
        & ${\rm d}\theta^{\rm rec}={\rm d}\theta^{\rm em}$
        \\
          $S_{\nu}^{\rm so}=\delta^{3}S_{\nu}^{\rm em}$
        & $S_{\nu}^{\rm rec}=S_{\nu}^{\rm so}/(1+z)^{3}$
        & $S_{\nu}^{\rm rec}=\delta^{3}S_{\nu}^{\rm em}/(1+z)^{3}$
        \\
        \hline
    \end{tabular}
    \tablefoot{
    The quantity on the emission frame ($Q^{\rm em}$) is affected by the Doppler effect, which in the source frame ($Q^{\rm so}$) is affected by redshift and in the receiver frame ($Q^{\rm rec}$) is affected by the Doppler effect and redshift.
    $\nu$ is the frequency, $A$ denotes the radiated surface area, $\theta$ is the angular size of the emission region, {and $S$ is the flux density.}
    $\delta$ is the Doppler factor, and $z$ is the redshift.
    }
\end{table} \\
\indent The linear scale ($R$) of the emission region is constrained by
\begin{equation}
    R_{\nu}^{\rm em}=gc\tau_{\nu}^{\rm em}=g\frac{c\delta\tau_{\nu}^{\rm rec}}{1+z},
    \label{eq:causality_linearsize}
\end{equation}
where $\nu$ is the observing frequency, $c$ is the speed of light, $\tau$ is a variability timescale, $\delta$ is the Doppler factor, and $g$ is a timescale scaling factor ($g=1$ for this work).
The angular size of the emission region at the receiver frequency $\nu^{\rm rec}$ measured by an observer in the receiver frame, $\theta_{R_{\nu}}^{\rm rec}$, is described as
\begin{equation}
    \theta_{R_{\nu}}^{\rm rec}=\frac{R_{\nu}^{\rm em}}{D_{{\rm A}}}=g\frac{c\delta\tau_{\nu}^{\rm rec}}{(1+z)D_{{\rm A}}}.
    \label{eq:causality_angularsize}
\end{equation}
The brightness temperature of the emission region $(T_{{\rm b}}^{\rm em})$ is defined in Rayleigh--Jeans regime as
\begin{equation}
    T_{{\rm b}}^{\rm em}=\frac{c^{2}}{2k_{{\rm {B}}}(\nu^{\rm em})^{2}}I_{\nu}^{\rm em},
    \label{eq:brightness_temperature}
\end{equation}
where $k_{{\rm B}}$ is the Boltzmann constant, and $I_{\nu}^{\rm em}$ is the intensity of the emission region at the frequency $\nu^{\rm em}$ in the emission frame.
The flux density $S_{\nu}$ is written as
\begin{equation}
    S_{\nu}
    =\int I_{\nu}(\theta, \phi){\rm d}\Omega
    =2\pi\int_{0}^{\theta_{R}}I_{\nu}(\theta)\sin\theta{\rm d}\theta
    =K\theta_{R}^{2}I_{\nu}(0),
    \label{eq:flux_to_intensity}
\end{equation}
where we assumed that $I_{\nu}(\theta, \phi)$ is azimuthally symmetric,  $\theta_{R}$ is the angular radius of the brightness distribution of the emission region (with a linear scale of $R$), and $K$ is a flux-scaling factor corresponding to the intensity distributions (Table~\ref{tab:scaling_factors}).
\begin{table}[!t]
    \caption{Scaling factors $K$ and $M$ that correct the distance}
    \label{tab:scaling_factors}
    \centering
    \begin{tabular}{llll}
        \hline\hline
        Morphology & $K$ & $M$ & $KM^{3}$ \\
        \hline
        Uniform Disk & $\pi$ & 0.7989 & 1.6019 \\
        Sphere & $2\pi/3$ & 0.9010 & 1.5319 \\
        \hline
    \end{tabular}
    \tablefoot{
    $K$ comes from Eq.~(\ref{eq:flux_density}). $KM^{3}$ is inversely proportional to the angular diameter distance (Eq.~(\ref{eq:distance_formula})).
    {$K$ and $M$ for each morphology are derived in Appendix~\ref{sect:A}.}
    }
\end{table}
The flux density is defined by
\begin{equation}
    S_{\nu}=\frac{h\nu{\rm d}^{3}n}{{\rm d}t{\rm d}A{\rm d}\nu},
    \label{eq:flux_density}
\end{equation}
where $n$ is the number of photons, and $h$ is the Planck constant.
The flux density in the emission frame is then obtained by~\citep{Boettcher2012}
\begin{equation}
    S_{\nu}^{\rm em}
    =\frac{\nu^{\rm em}}{\nu^{\rm rec}}
    \frac{{\rm d}t^{\rm rec}}{{\rm d}t^{\rm em}}
    \frac{{\rm d}A^{\rm rec}}{{\rm d}A^{\rm em}}
    \frac{{\rm d}\nu^{\rm rec}}{{\rm d}\nu^{\rm em}}
    S_{\nu}^{\rm rec}
    =\frac{(1+z)^{3}}{\delta^3}S_{\nu}^{\rm rec}.
    \label{eq:flux_frame_transform}
\end{equation}
Hence, using Eqs.~(\ref{eq:brightness_temperature}), (\ref{eq:flux_to_intensity}), and (\ref{eq:flux_frame_transform}), and Table~\ref{tab:frame_transform}, the brightness temperature in the emission frame can be written as
\begin{equation}
    T_{{\rm b}}^{\rm em}
    =\frac{c^{2}}{2k_{{\rm {B}}}{\nu^{\rm rec}}^{2}}
    \frac{S_{\nu}^{\rm rec}}{K{\theta_{R_{\nu}}^{\rm rec}}^{2}}
    \frac{1+z}{\delta}.
    \label{eq:brightness_temperature_transform}
\end{equation}
Depending on whether the angular size of the emission region is defined using a variability timescale or a VLBI image, the brightness temperature can be divided into a variability brightness temperature and VLBI brightness temperature.
The variability brightness temperature in the emission frame ($T_{{\rm b,var}}^{\rm em}$) is defined when $\theta_{R_{\nu}}^{\rm rec}$ is measured by the size--timescale causality (Eq.~(\ref{eq:causality_angularsize})) as
\begin{equation}
    T_{{\rm b,var}}^{\rm em}
    =\frac{1}{2k_{{\rm {B}}}{\nu^{\rm rec}}^{2}}
    \frac{S_{\nu}^{\rm rec}{D_{{\rm A}}}^{2}}{K(g\tau_{\nu}^{\rm rec})^{2}}
    \frac{(1+z)^{3}}{\delta^{3}}.
    \label{eq:variability_brightness_temperature}
\end{equation}
The VLBI brightness temperature in the emission frame ($T_{{\rm b,VLBI}}^{\rm em}$) is defined when $\theta_{R_{\nu}}^{\rm rec}=M\theta_{{\rm FWHM},\nu}$ (Table~\ref{tab:scaling_factors}) as
\begin{equation}
    T_{{\rm b,VLBI}}^{\rm em}
    =\frac{c^2}{2k_{\rm B}{\nu^{\rm rec}}^{2}}
    \frac{S_{\nu}^{\rm rec}}{K(M\theta_{{\rm FWHM},\nu}^{\rm rec})^{2}}
    \frac{1+z}{\delta},
    \label{eq:VLBI_brightness_temperature}
\end{equation}
where $\theta_{{\rm FWHM},\nu}$ is the angular full width at half maximum (FWHM) of a circular Gaussian model component in VLBI images, and $M$ is the size scaling factor.
We introduce the intrinsic brightness temperature $T_{\rm b,int}$ which constrains the maximum brightness temperature of AGNs in the emission region.
We assumed that $T_{{\rm b,var}}^{\rm em}$ and $T_{{\rm b,VLBI}}^{\rm em}$ are equal to $T_{\rm b,int}$.
By dividing the cube of Eq.~(\ref{eq:VLBI_brightness_temperature}) by Eq.~(\ref{eq:variability_brightness_temperature}) and then taking the square root, $T_{\rm b,int}$ can be expressed as
\begin{equation}
    T_{\rm b,int}
    =\sqrt{\frac{{T_{{\rm b,VLBI}}^{\rm em}}^{3}}{T_{\rm b,var}^{\rm em}}}
    =\frac{c^{3}}{2k_{\rm B}{\nu^{\rm rec}}^{2}}\frac{S_{\nu}^{\rm rec}g\tau_{\nu}^{\rm rec}}{K(M\theta_{{\rm FWHM},\nu}^{\rm rec})^{3}D_{{\rm A}}}.
    \label{eq:intrinsic_brightness_temperature}
\end{equation}
Finally, the angular diameter distance is determined as
\begin{equation}
    D_{A}
    =\frac{c^{3}}{2k_{{\rm B}}T_{\rm b,int}{\nu^{\rm rec}}^{2}}
    \frac{S_{\nu}^{\rm rec}g\tau_{\nu}^{\rm rec}}{K(M\theta_{{\rm FWHM},\nu}^{\rm rec})^{3}},
    \label{eq:distance_formula}
\end{equation}
which is a revised form of the formula by~\citet{Hodgson2023}.

\section{Data analysis}\label{sect:3}
To measure the angular sizes of the emission regions and the timescales and peak flux densities of flares, we used 32 epochs of 15~GHz Very Long Baseline Array (VLBA) data from MJD~54838 to MJD~60262 (15~years), which is part of the Monitoring Of Jets in Active galactic nuclei with VLBA Experiments (MOJAVE;~\citep{Lister2018}).
For more precise timescales and peak flux densities, we additionally used 620 epochs of 15~GHz Owens Valley Radio Observatory (OVRO) 40~m single-dish (SD) data from MJD~54474 to MJD~59023 (12~years; \citep{Richards2011}, which have a mean cadence of about 7~days (significantly denser than the $\sim170$-day mean cadence of the VLBA data).

\subsection{Circular Gaussian fitting to 15~GHz VLBA data from MOJAVE}\label{sect:3.1}
\begin{figure}[!t]
    \centering
    \includegraphics[width=1.0\linewidth]{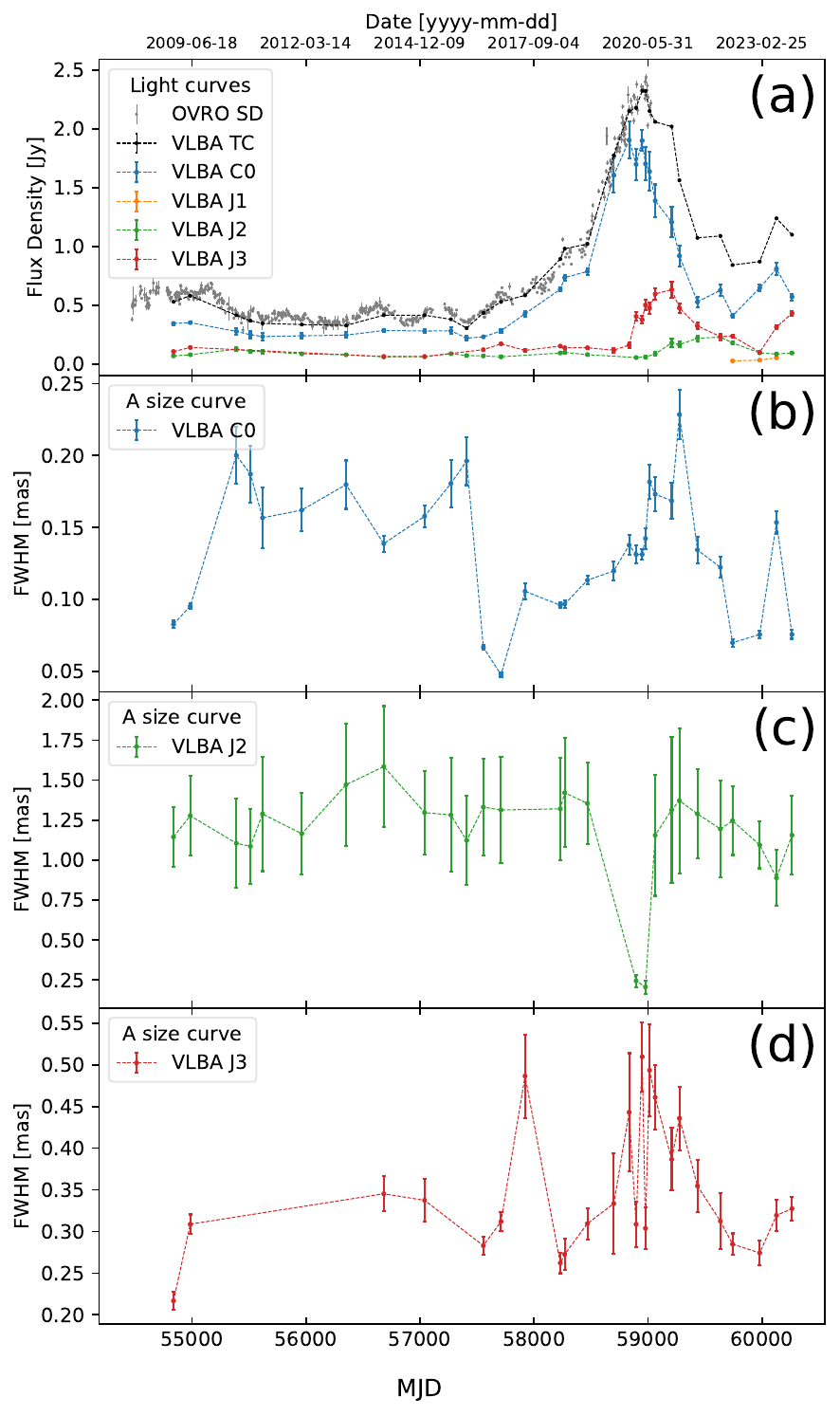}
    \caption{
    Panel~(a): Flux density curves for each VLBA circular Gaussian model component (colored dashed lines), VLBA total clean (dashed black line), and OVRO~SD (gray dots). 
    Panels~(b), (c), and (d): FWHM curves for VLBA~C0, J2, and J3.
    The VLBA model components are defined by their positions as in Fig.~\ref{fig:positions}.
    }
    \label{fig:light_size_curves}
\end{figure}
\begin{figure}[!t]
    \centering
    \includegraphics[width=1.0\linewidth]{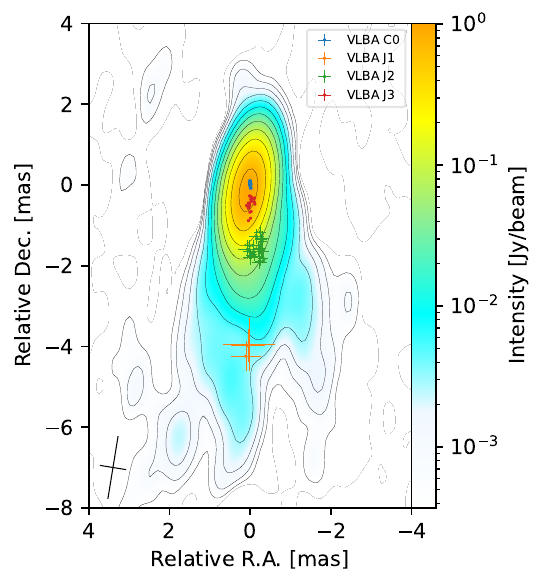}
    \caption{
    Clean map of TXS~0506+056 on MJD~60126 (July~1, 2023) from MOJAVE overlapped with multiepoch positions of Gaussian model components (VLBA~C0, J1, J2, and J3).
    The major and minor FWHMs of the elliptical clean beam are plotted in the lower left corner.
    The contours are given at  $\log_{2}$ level from three~times the root mean square to the peak intensity.
    }
    \label{fig:positions}
\end{figure}
To measure the variability timescales and the corresponding angular sizes of the emission region, we fit the 2D circular Gaussian model to the 15~GHz VLBA data using the program \texttt{DIFMAP}~\citep{Shepherd1994}, as described in Appendix~\ref{sect:B}.
The total results of the Gaussian model fitting are presented in Table~\ref{tab:VLBA_model_fitting_result}.
We defined the core component, which is the nearest to the center of images, as C0 and other jet components as J1, J2, and J3 (Fig.~\ref{fig:positions}).
We found that the source has a core--jet structure in which the core flux density ranges from 0.221~Jy to 1.906~Jy and the jet flux densities range from 0.073--0.814~Jy. This yields a total model flux density (core and jet flux densities) of 0.294--2.278~Jy.
Fig.~\ref{fig:light_size_curves} shows that at the core (VLBA~C0), the largest flare peaks on MJD~58834 and the two jet components (VLBA~J2 and J3) have a higher flux with a peak on MJD~59433 and MJD~59275, respectively.
We also found that the core size changes in the range of 0.048--0.228~mas.

\subsection{Flare decomposition of the 15~GHz VLBA model components and OVRO 40~m SD light curves}\label{sect:3.2}
\begin{figure}[!ht]
    \centering
    \includegraphics[width=1\linewidth]{fig/flare_decomposition.pdf}
    \caption{
    Flare decomposition plots of the VLBA~C0~(a), J2~(b), and J3~(c), respectively.
    For a purpose of comparison, the flare model components in panels~(a)--(c) overlap the OVRO~SD+VLBA~TC light curve in panel~(d).
    A flare decomposition plot (e) of the OVRO~SD+VLBA~TC light curve using the initial sample shown in panel (d).
    The solid lines are the single flare model components.
    The dashed line in each panel describes the sum of all flare model components and the quiescent flux density.
    All panels note the number of components $N$, log-likelihood $\log \mathcal{L}$, $\chi^{2}$, and reduced $\chi^{2}_{d}$ ($d$ is the degrees of freedom) in the upper left corner.
    The quiescent flux density ($F_{\rm qs}$) is noted in the lower left corner (as well as by the horizontal dotted lines).
    }
    \label{fig:decomposition}
\end{figure}
We decomposed the VLBA light curves using the diffusive nested sampling Python package \texttt{dnest4}~\citep{Brewer2018}, as described in Appendix~\ref{sect:C}.
The results are shown in Fig.~\ref{fig:decomposition} and are summarized in Table~\ref{tab:flare_decomposition_parameters}.
In addition, we performed the same decomposition process using the OVRO~SD light curve
by imposing the decomposition results of the VLBA~C0, J2 and J3 as initial parameters.
The OVRO data end at MJD~59023, however, while the VLBA data extend to MJD~60262.
Therefore, the OVRO~SD flux density after MJD~59023 was replaced by the VLBA total clean (VLBA~TC) flux density from MJD~59062 to MJD~{60262}.
We named the joint light curve between OVRO~SD (MJD~54474--59023) and VLBA TC (MJD~59062--{60262}) OVRO~SD+VLBA+TC.

\section{Results}\label{sect:4}
\subsection{Comparison between cross-identified flares}\label{sect:4.1}
\begin{table}[!ht]
    \caption{Cross-identified flares}
    \label{tab:cross_identified_flares}
    \centering
    \begin{tabular}{ll@{\extracolsep{10pt}}l@{\extracolsep{8pt}}l@{\extracolsep{8pt}}l@{\extracolsep{8pt}}l}
        \hline\hline
        \multicolumn{2}{c}{Identified flare} & \multicolumn{3}{c}{Fractional offset} & ${\rm NED}^{2}$ \\
        \cline{1-2}\cline{3-5}
        VLBA & SD & $\Delta\tau$ & $\Delta F_{\rm p}$ & $\Delta(\tau F_{\rm p})$ & ~ \\
        \hline
        C0a & SD4 & $-0.108$ & $-0.010$ & $-0.117$ & 0.0259 \\
        C0c & SD14 & $-0.132$ & $-0.223$ & $-0.326$ & 0.0804 \\
        C0d & SD19 & $0.026$ & $0.090$ & $0.119$ & 0.0245 \\
        C0e & SD24 & $-0.163$ & $-0.129$ & $-0.271$ & 0.0238 \\
        \hline
    \end{tabular}
    \tablefoot{
    $\Delta\tau=(\tau_{i}-\tau_{j})/\tau_{j}$, $\Delta F_{\rm p}=(F_{{\rm p},i}-F_{{\rm p},j})/F_{{\rm p},j}$, $\Delta(\tau F_{\rm p})=(\tau_{i}F_{{\rm p},i}-\tau_{j}F_{{\rm p},j})/(\tau_{j}F_{{\rm p},j})$, where $i$ and $j$ are cross-identified flare indices from the OVRO~SD+VLBA~TC and VLBA light curves, respectively. ${\rm NED}^{2}$ is the normalized squared Euclidean distance described in Eq.~(\ref{eq:normalized_squared_Euclidean_distance}), which quantifies the similarity between flare models $F_{i}(t)$ and $F_{j}(t)$.
    }
\end{table}
\begin{figure*}[!ht]
    \centering
    \includegraphics[width=1.0\linewidth]{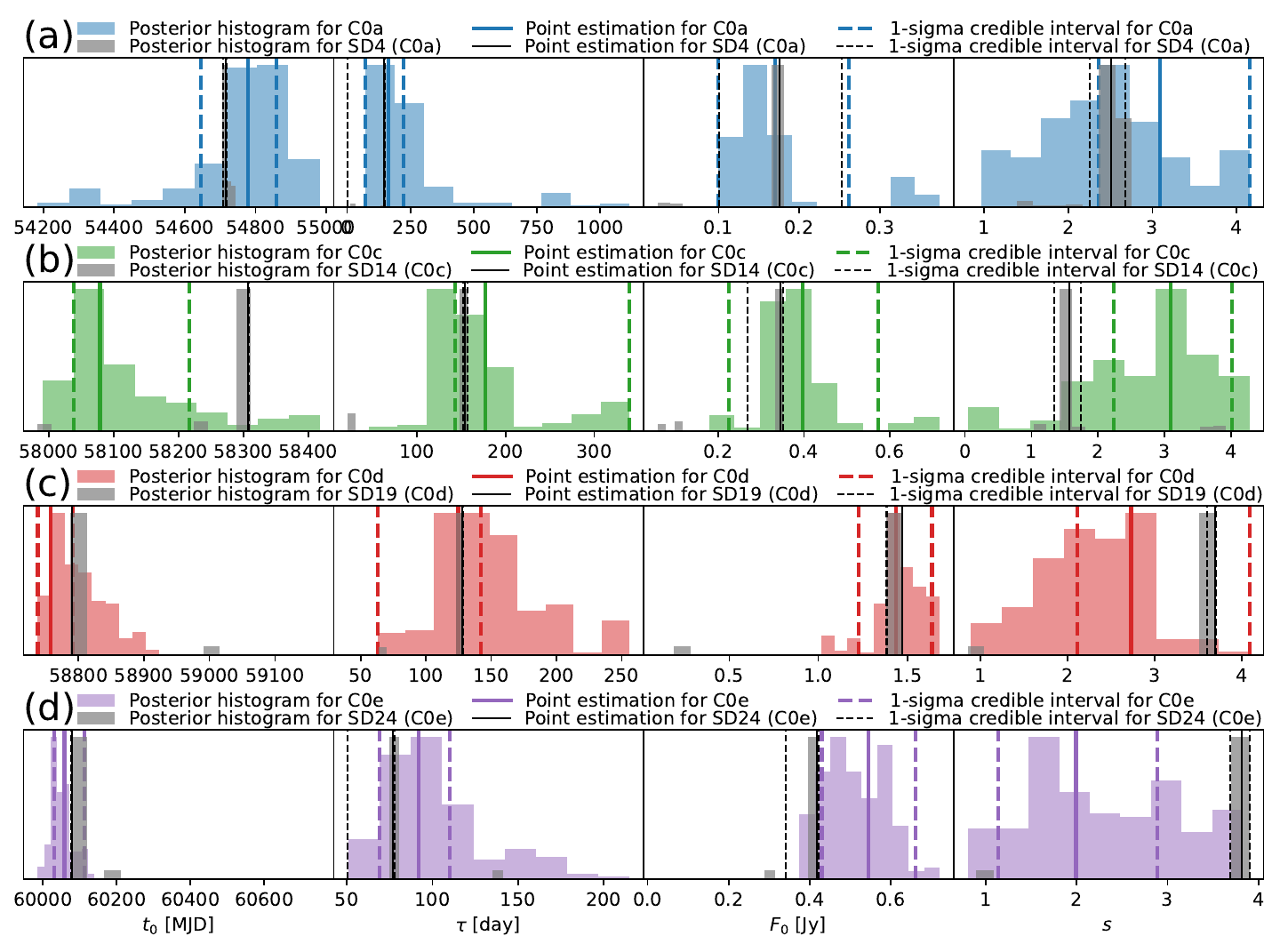}
    \caption{
    Posterior distributions, point estimates, and credible intervals of the cross-identified flare decomposition parameters $t_{0}$, $\tau$, $F_{0}$, and $s$ for flares C0a, C0c, C0d, and C0e (color plots from panels~(a) to (d)) and SD4~(C0a), SD14~(C0c), SD19~(C0d), and SD24~(C0e) (gray and black plots from panels~(a) to (d)).
    The bar plots illustrate the posterior distributions.
    The solid and dashed line plots note the point estimates and interval estimates, respectively.
    }
    \label{fig:posterior_distributions}
\end{figure*}
We found that some of the flares decomposed from the VLBA light curves are identified in the OVRO~SD+VLBA~TC light curve, as noted in Table~\ref{tab:flare_decomposition_parameters}.
We considered the flares from OVRO~SD+VLBA~TC as cross-identifiers whose parameter estimates (reference time $t_{0}$, rising timescale $\tau$, reference flux density $F_{0}$, and skewness $s$) are uniquely identified in the range of their posterior samples from the VLBA light curves, that is, VLBA C0a, C0c, C0d, and C0e (see Fig.~\ref{fig:posterior_distributions}).
The decomposition parameters of the cross-identified flares are compared in Table~\ref{tab:cross_identified_flares}. This comparison shows that the fractional difference of the timescale is $|\Delta\tau|=0.026$--$0.163$ and that of the peak flux density is $|\Delta F_{\rm p}|=0.010$--$0.223$.
The fractional offset of distances ($D_{\rm A}$) between the cross-identifiers is similar to that of $\tau\times F_{\rm p}$ (c.f., $D_{\rm A}\propto\tau F_{\rm p}$ and Eq.~(\ref{eq:distance_formula})), yielding $|\Delta(\tau F_{\rm p})|=0.117$--$0.326$.
The similarity between the cross-identified flares was also evaluated using the normalized squared Euclidean distance (${\rm NED}^{2}_{ij}$), which is defined as in Eq.~(\ref{eq:normalized_squared_Euclidean_distance}), where Var is the variance, and $F_{i}(t)$ is the flare model function described in Eq.~(\ref{eq:flare_model_function}),
\begin{equation}
    {\rm NED}^{2}_{ij}=0.5\frac{{\rm Var}(F_{i}(t)-F_{j}(t))}{{\rm Var}(F_{i}(t))+{\rm Var}(F_{j}(t))}
    \label{eq:normalized_squared_Euclidean_distance}.
\end{equation}
${\rm NED}^{2}$ is equal to zero or far lower than unity for an identical flare, and it approaches unity for significantly different flares.
${\rm NED}^{2}$ is found to be in the range of 0.0238--0.0804, which is consistent with it being an identical flare.

\subsection{Constraining the distance $(D_{\rm A})$}\label{sect:4.2}
\begin{figure}[!t]
    \centering
    \includegraphics[width=1\linewidth]{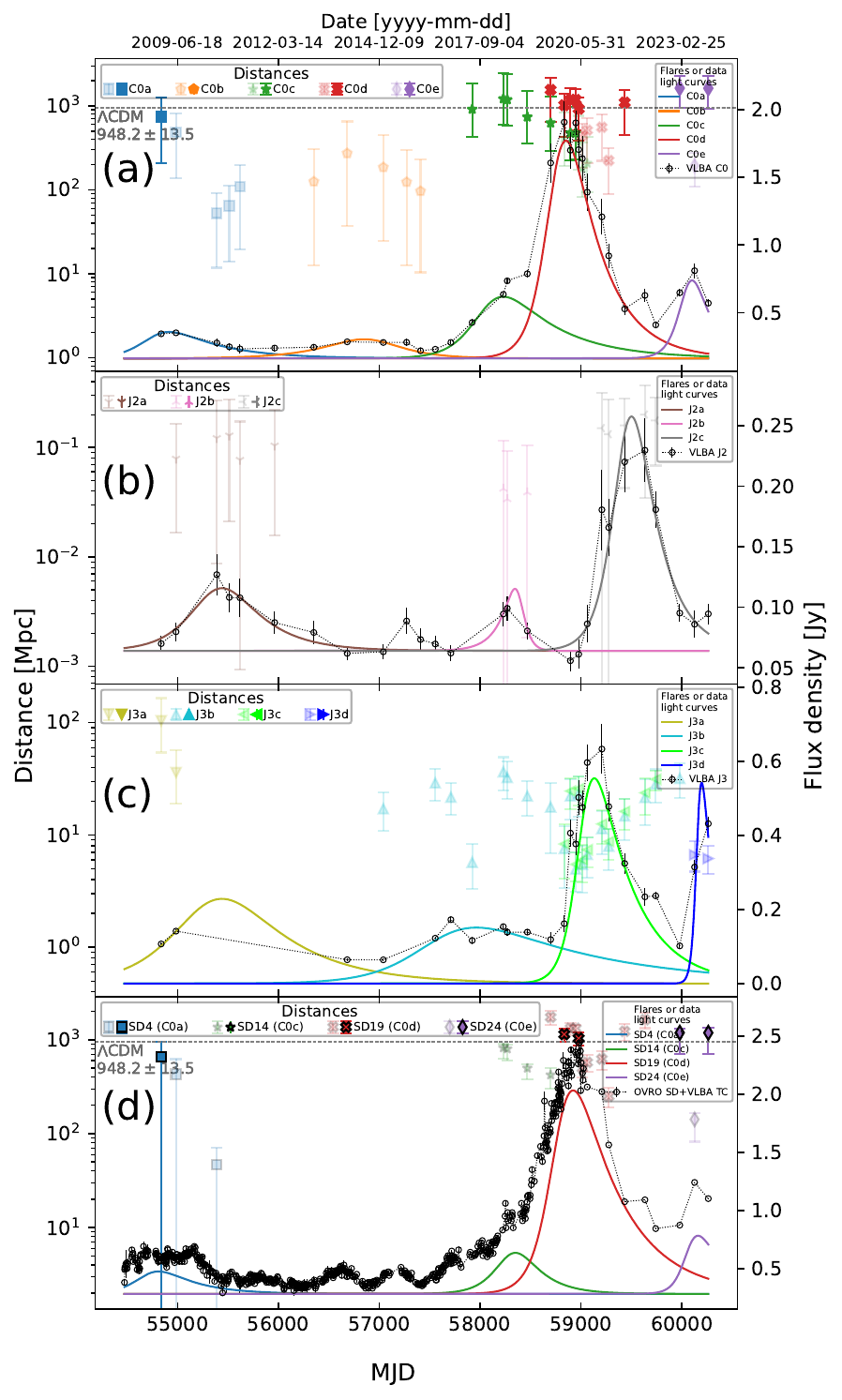}
    \caption{
    Distance estimates of VLBA~C0, J2, J3, and OVRO~SD+VLBA~TC (from panels~(a) to (d)). The solid colored lines show decomposed flares, the colored symbols show the distance estimates using timescales and peak flux densities from the individual flares in corresponding colors, and the black circles show the data (VLBA and OVRO) light curves.
    The horizontal dashed gray lines note the $\Lambda$CDM distance ($948.2\pm13.5$~Mpc) with 1$\sigma$ uncertainty.
    The color symbols of the distance estimates are opaque when the distance measurements are consistent with the $\Lambda$CDM distance within 1$\sigma$ uncertainties, and they are transparent when the distances are not.
    }
    \label{fig:distances}
\end{figure}
We determined that the flux scaling factor is $K=\pi$ and the size scaling factor is $M=0.8$ (see Table~\ref{tab:scaling_factors}).
We assumed that the timescale measures the radius perfectly. The timescale scaling factor then is $g=1$.
We used the equipartition temperature $T_{\rm eq}=5\times10^{10}$~K~\citep{Readhead1994} as the intrinsic brightness temperature.
We investigated the distance estimates using the core sizes obtained around flares in the period from $t_{0}-2\tau$ to $t_{0}+2s\tau$ for each flare with its timescale $\tau$ and peak flux density $F_{\rm p}$, as summarized in Tables~\ref{tab:distances_C0+SD} and \ref{tab:distances_J2+J3}.
The results of the distance measurements are shown in Fig.~\ref{fig:distances}, and their 1$\sigma$ errors were calculated with the Python package \texttt{uncertainties}.
Fig.~\ref{fig:distances} shows transparent symbols for the distance measurements that deviate by more than 1$\sigma$ uncertainties from $D_{A}=948.2\pm13.5$~Mpc, which was calculated with the $\Lambda$CDM model, where $H_{0}=73.04\pm1.04$~km/s/Mpc and $\Omega_{\rm m}=0.315$.
When we used the timescale and peak flux density from the VLBA data, all flares of VLBA~C0 yield distance estimates that are consistent with the $\Lambda$CDM, while those of VLBA~J2 and J3 are lower than the $\Lambda$CDM distance by 2--4 orders of magnitude.
To improve the statistical errors, we also used flares SD4, SD14, SD19, and SD24 from the OVRO~SD+VLBA~TC data, which were cross-identified with flares C0a, C0c, C0d, and C0e from the VLBA data.
Consistent distance measurements were largely obtained using parameters near the flare peak time (see Sect.~\ref{sect:5.3} for further discussion).

\section{Discussions}\label{sect:5}
\subsection{Error recognition}\label{sect:5.1}
The fractional uncertainties of distances for the cross identifiers (C0d/C0e and SD19/SD24) are 40--50~\% for the VLBA~C0 data and 15--26~\% for the OVRO~SD+VLBA~TC data.
That the uncertainties for OVRO~SD+VLBA~TC improve by factors of 2--3 is attributed to the fact that the data cadence is about 18.4 times higher and their root mean square error of the flux density is lower by about 3.6 times than that of the VLBA data.
The high cadence and precise flux measurements mean that the fractional uncertainties of the distances are smaller by $<5$ times for the cross identifiers.
The OVRO~SD+VLBA~TC light curve is the total flux density of the core and all jet components, while the VLBA light curves correspond to individual components.
To avoid a potential systematic error from the blending effect of the core and jet components, we decomposed the OVRO~SD+VLBA~TC light curve as described in Sect.~\ref{sect:3.2} and Appendix~\ref{sect:C}.
To examine the effect of the systematic error, we compared the parameter offsets (the timescale and peak flux density) between the VLBA and OVRO~SD+VLBA~TC with the corresponding parameter uncertainties of the VLBA~C0 flares (C0a, C0c, C0d, and C0e), as shown in Fig.~\ref{fig:fractional_errors}.
We found that the offsets in the flares are within the corresponding parameter uncertainties of VLBA~C0.

\subsection{Comparison with the standard e-folding timescale}\label{sect:5.2}
\begin{figure}[!t]
    \centering
    \includegraphics[width=1\linewidth]{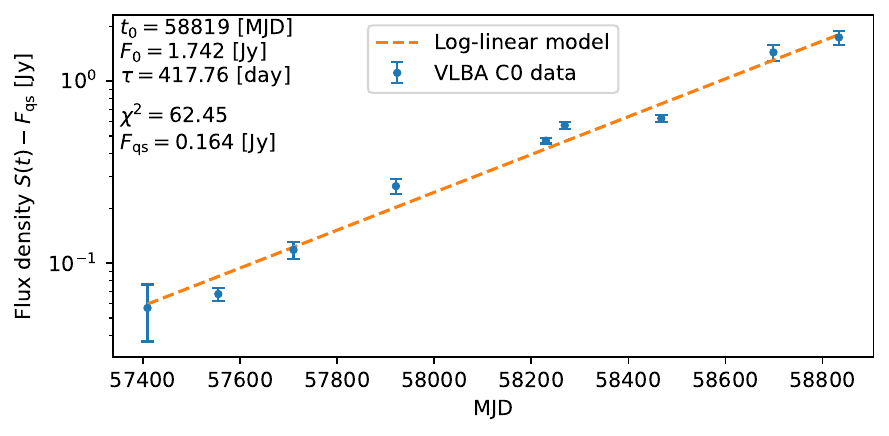}
    \caption{
    Log-linear fit to the VLBA~C0 light curve from MJD~58230 to MJD~58895 and given quiescent flux density $F_{\rm qs}=0.164$~Jy.
    The fitting function is $S(t)-F_{\rm qs}=F_{0}\exp\big(\frac{t-t_{0}}{\tau}\big)$.
    The reference time $t_{0}$, the reference flux density $F_{0}$, the standard e-folding timescale $\tau$, $\chi^{2}$, and the quiescent flux density $F_{\rm qs}$ are noted in the upper right corner.
    }
    \label{fig:standard_efolding}
\end{figure}
Characteristic timescales of variable emission regions in blazars are determined with various methods, including a log-linear fit (a standard e-folding timescale), a flare decomposition (used in this paper), and a structure function calculation.
While the structure function calculation is strongly limited by the data cadence in timescale resolution, the log-linear fit and flare decomposition are free from this limitation.
The former two methods are therefore probably largely used to determine the characteristic variability timescale of blazars \citep[e.g.,][]{Jorstad2017, Liodakis2017}.
The log-linear fit was used to characterize the variability timescales of resolved jet components \citep[e.g.,][]{Jorstad2017, Hodgson2020}, and the flare decomposition method was largely used to decompose the individual flares in the light curves obtained from single-dish radio observations \citep[e.g.,][]{Liodakis2017, Liodakis2018, Kang2021}.\\
\indent In order to compare the timescales of two methods, we fit a log-linear function to the VLBA~C0 light curve for the time period of MJD~57409--58834 (Fig.~\ref{fig:standard_efolding}), which yielded an e-folding timescale of about 418~days. We compared this with the decomposition timescales of 178~days (C0c) and 125~days (C0d).
The difference of the timescales between the standard e-folding and decomposition cases is most likely attributable to the overlapping multiple flares C0c (peaking on MJD~58230 and reaching $F_{\rm p}=0.46$~Jy) and C0d (peaking on MJD~58850 and reaching
$F_{\rm p}=1.61$~Jy).
The standard e-folding timescale for the VLBA~C0 light curve is similar to the peak-to-peak e-folding timescale for the flares C0c and C0d (about 495~days).
Then, the standard e-folding timescale is affected by the flare overlapping and hence overestimated.
In order to compare the e-folding timescale of a selected period in individual decomposed flares
(e.g., VLBA C0) with the corresponding timescales of a decomposed flare, we investigated the model function and its derivative, as summarized in Appendix~\ref{sect:F}.
We found that the e-folding timescale becomes longer than the decomposed timescale, in particular, at the peak of the flare.
Therefore, we conclude that the decomposition timescales derived in this study are more robust than the standard e-folding timescales, owing to mitigating the systematic errors of flare overlapping and selecting fitting period.

\subsection{Distance determination from blazars}\label{sect:5.3}
\begin{figure}[!t]
    \centering
    \includegraphics[width=1\linewidth]{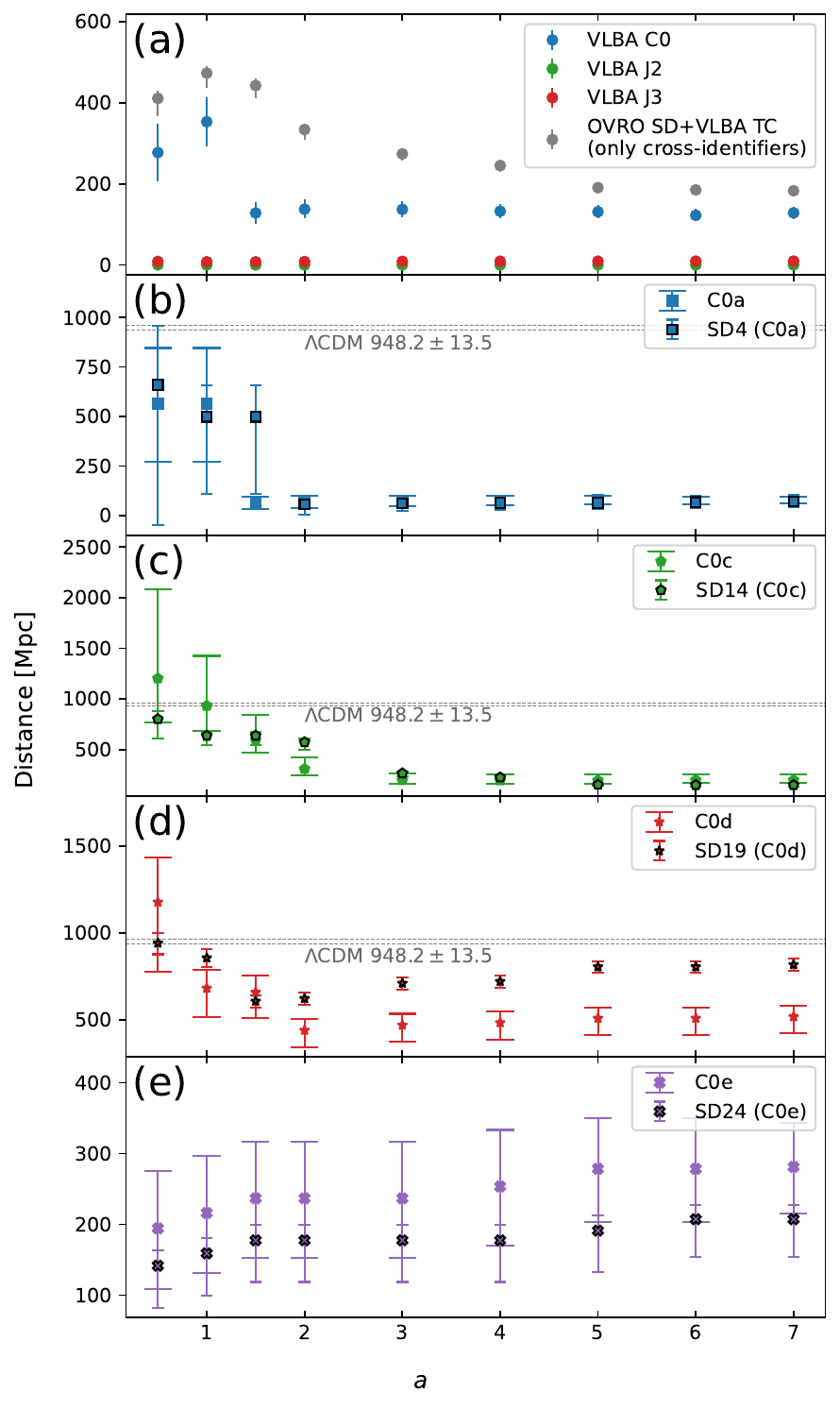}
    \caption{
    {Weighted mean of distances. The distances were calculated using the timescales and peak flux densities of each flare (C0a, C0b, C0c, C0d, C0e, J2a, J2b, J2c, J3a, J3b, J3c, and J3d) and the angular sizes of the corresponding individual components (VLBA~C0, J2, and J3)} obtained in the time from $t_{0}-a\tau$ to $t_{0}+as\tau$.
    Panel~(a) shows the weighted means of all distances estimated for VLBA~C0 (blue circles), J2 (green circles), J3 (red circles), and OVRO~SD+VLBA~TC (gray circles).
    The other panels~(b)--(e) show the weighted mean distances of C0a (blue squares), C0c (green pentagons), C0d (red stars), and C0e (purple crosses), respectively, with the weighted mean distances for the corresponding cross-identifiers (black outlined symbols).
    The horizontal dashed gray lines show the $\Lambda$CDM distance ($948.2\pm13.5$~Mpc) with the 1$\sigma$ uncertainty.
    }
    \label{fig:weighted_mean_distances}
\end{figure}

To determine the appropriate angular sizes of the emission regions, we investigated the systematics of the angular sizes on the distance estimates.
We calculated the weighted mean distance using the timescales and peak flux densities of each flare (C0a, C0b, C0c, C0d, C0e, J2a, J2b, J2c, J3a, J3b, J3c, and J3d) and the angular sizes of the corresponding individual components (VLBA~C0, J2, and J3) obtained in the time from $t_{0}-a\tau$ to $t_{0}+as\tau$, as described in Fig.~\ref{fig:weighted_mean_distances}.
As the time range of averaging distances increases (i.e., $a$ increases), the weighted mean distance largely decreases (i.e., a larger offset to the $\Lambda$CDM distance, except for C0e and SD24), although the corresponding scatter uncertainty decreases.
To determine accurate distances, we chose angular sizes near flare peaks.
All of the cross-identified flares SD4~(C0a), SD14~(C0c), SD19~(C0d), and SD24~(C0e) yielded consistent distance measurements using the accurate timescales and peak flux densities from the high-cadence SD light curve.
The C0b flare was not cross-identified with any of the SD flares, and it therefore yielded inaccurate distance measurements.
Additionally, we found that the largest flare SD19~(C0d) enabled us to obtain the most precise and accurate distance of the source.

\subsection{Unresolvable components}\label{sect:5.4}
In a Gaussian fitting, we were unable to obtain a solution that converged to only one for each epoch because of the unresolvable but apparently significant components near the core.
In the period of MJD~54838--58468, the reduced $\chi^{2}$ of the best-fitting Gaussian models was improved to 1.071--1.636.
From MJD~58699 to MJD~59207, however, the reduced $\chi^{2}$ of the best-fitting Gaussian model increased to 5.734--9.546.
The errors of the core flux densities and FWHM are therefore larger because of the high $\sigma_{\rm rms}$ (Eq.~(\ref{eq:Gaussian_model_error})).
When unresolvable components are added near the core, the $\chi^{2}$ might be improved to about 1--2.
Because the unresolvable component does not appear continuously in the fitting, however, the flux of the core looks as if it fluctuated. This prevents us from determining an accurate timescale and flux density variation in VLBA~C0.
If the VLBI observation had a high resolution and dynamic range, they would be resolvable.

\section{Conclusions}\label{sect:6}
We measured the distances of blazar TXS~0506+056 using the variability timescale, the flare peak flux density, and the angular sizes of the core and jets, and we assumed an equipartition brightness temperature.
The variability timescale and the flare peak flux density were obtained from the 15~GHz VLBA and OVRO~SD+VLBA~TC light curves.
The core sizes were measured using the 15~GHz VLBA data.
We summarize our findings below.
\begin{enumerate}
\item We obtained distance measurements consistent with the $\Lambda$CDM model within a 1$\sigma$ uncertainty when the core sizes were measured in the period of $t_{0}-2\tau$ to $t_{0}+2s\tau$, which includes the flare peaks for each flare.
This implies that the optimal angular size of the flaring core is well determined near the flare peak for measuring a consistent distance.
\item  We also emphasize that the distance estimates are underestimated when the angular sizes of the jet components are used in any time period.
This implies that the distance estimations seem to work well for the core component, probably because the angular sizes of the core are properly measured, whereas those of jet are overestimated.
It is necessary to investigate this issue further in future studies.
\item The high cadence and small errors of the OVRO data meant that the combined data of OVRO~SD and VLBA~TC data enabled a far preciser estimation than the VLBA~C0 data.
\item After avoiding a potential systematic error from the blending effect of the core and jet components with the OVRO SD+VLBA TC light-curve decomposition, we found that the offsets in the flares are within the corresponding parameter uncertainties of VLBA~C0.
\item One of the dominant factors affecting the distance measurements is $\theta_{\rm FWHM}$ because the distance depends on $\theta_{{\rm FWHM}}^{-3}$, whose uncertainty is as large as 40~\% for VLBA~C0.
\item We found that the decomposed timescale enabled us to obtain distances that were consistent with the $\Lambda$CDM.
We strongly suggest to decompose the light curves when the variability timescales are to be obtained properly.
\item In a circular Gaussian fitting, we decided to fit only resolvable components.
From MJD~58699 to MJD~59207, we were only able to improve the $\chi^{2}$ in a range of 5.734--9.546, however, except for unresolvable components.
When more high-resolution VLBI observations over a wider dynamic range are available, preciser model parameters will be obtained, such as $\theta_{\rm FWHM}$, and the uncertainty of distance measurement can be decreased.
\item We suggest that the best distance of the source is $941_{-64}^{+59}$~Mpc, that is, the weighted mean of distances from $t_{0}-0.5\tau$ to $t_{0}+0.5s\tau$ of the largest OVRO~SD flare (SD19). This is comparable to the $\Lambda$CDM distance of $948.2\pm13.5$~Mpc.
The results were obtained under the assumption of a  disk brightness morphology, an equipartition brightness temperature ($T_{\rm b,int}=5\times10^{10}$~K), and a perfect radius ($g=1$).
\end{enumerate}

\begin{acknowledgements}
This work was supported by the National Research Foundation of Korea (NRF) grant funded by the Korea government (MIST) (2020R1A2C2009003, RS-2025-00562700).
This research has made use of data from the MOJAVE database that is maintained by the MOJAVE team (Lister et al. 2018).
This research has made use of data from the OVRO 40-m monitoring program (Richards, J. L. et al. 2011, ApJS, 194, 29), supported by private funding from the California Institute of Technology and the Max Planck Institute for Radio Astronomy, and by NASA grants NNX08AW31G, NNX11A043G, and NNX14AQ89G and NSF grants AST-0808050 and AST- 1109911.
Uncertainties: a Python package for calculations with uncertainties, Eric O. LEBIGOT, http://pythonhosted.org/uncertainties/.
\end{acknowledgements}

\bibpunct{(}{)}{;}{a}{}{,}
\bibliographystyle{bibtex/aa}
\bibliography{bibtex/references.bib}

\begin{appendix}
\section{Size and flux scaling along projected brightness morphologies}\label{sect:A}
\begin{table}[!ht]
    \caption{Intensities of elliptic cylindrical symmetric morphologies.}
    \label{tab:morphologies_intensity}
    \centering
    \begin{tabular}{ll}
        \hline\hline
        Morphology & Intensity $I(p)$\\
        \hline
        Disk & $I_{0}H(1-p)$ \\
        Ellipsoid & $I_{0}\sqrt{1-p^{2}}H(1-p)$ \\
        Gaussian & $I_{0}\exp{\Big(-\frac{s^{2}p^{2}}{2}\Big)}$ \\
        \hline
    \end{tabular}
    \tablefoot{
    $H(x)$ is the Heaviside function, $I_{0}$ is the center intensity where $p=0$, $p$ is the normalized coordinate defined as Eq.~(\ref{eq:normalized_intensity_coordinate}), and $s$ is the factor defining the Gaussian morphology as $\theta_{R}=s\sigma_{I}$ in Eq.~(\ref{eq:normalized_intensity_coordinate}), where $\sigma_{I}$ is the standard deviation of the Gaussian intensity distribution.
    }
\end{table}
Suppose that the projected brightness morphology of the emission region is elliptic cylindrical symmetric.
The intensities of example morphologies are shown in Table~\ref{tab:morphologies_intensity}.
$p$ is the normalized coordinate defined as
\begin{equation}
    p\equiv\frac{1}{\theta_{R}}\sqrt{\frac{l^{2}}{a^{2}}+m^{2}}=\frac{r}{\theta_{R}}\sqrt{\frac{\sin^{2}{\phi}}{a^{2}}+\cos^{2}{\phi}},
    \label{eq:normalized_intensity_coordinate}
\end{equation}
where $(l,m)$ is the equatorial coordinate system, and $(r,\phi)$ is the corresponding polar coordinate system.
$a$ is the ratio of the minor to major axes $(0<a\leq1)$.
$\theta_{R}$ is the radius of major axis for finite morphologies such as disk, ellipsoid, and cone.
$\theta_{R}$ is the $s$-$\sigma$ boundary of major axis for a Gaussian morphology whose size is infinite.
When $s=\sqrt{2\ln{2}}$ for a Gaussian morphology, $\theta_{R}=\theta_{\rm FWHM}/2$.
$I_{0}$ is the center intensity, where $p=0$.
The flux density ($S$) of morphologies is calculated as
\begin{equation}
    \begin{split}
        S&=\int_{0}^{2\pi}\int_{0}^{\infty}I(r,\phi)r{\rm d}r{\rm d}\phi \\
        &=\int_{0}^{2\pi}\int_{0}^{\infty}\theta_{R}^{2}\bigg(\frac{\sin^{2}{\phi}}{a^{2}}+\cos^{2}{\phi}\bigg)^{-1}I(p)p{\rm d}p{\rm d}\phi \\
        &=\theta_{R}^{2}\int_{0}^{2\pi}\bigg(\frac{\sin^{2}{\phi}}{a^{2}}+\cos^{2}{\phi}\bigg)^{-1}{\rm d}\phi\int_{0}^{\infty}I(p)p{\rm d}p \\
        &=2\pi a\theta_{R}^{2}\int_{0}^{\infty}I(p)p{\rm d}p=Ka\theta_{R}^{2}I_{0},
    \end{split}
    \label{eq:morphology_flux_density}
\end{equation}
where the flux scaling factor $K$ is defined as
\begin{equation}
    K\equiv\frac{2\pi}{I_{0}}\int_{0}^{\infty}I(p)p{\rm d}p.
    \label{eq:flux_scaling_factor}
\end{equation}
Visibility is Fourier transformation of the intensity, and the visibility functions of the intensity distributions in Table~\ref{tab:morphologies_intensity} are shown in Table~\ref{tab:morphologies_visibility}.
\begin{figure}[!t]
    \includegraphics[width=1\linewidth]{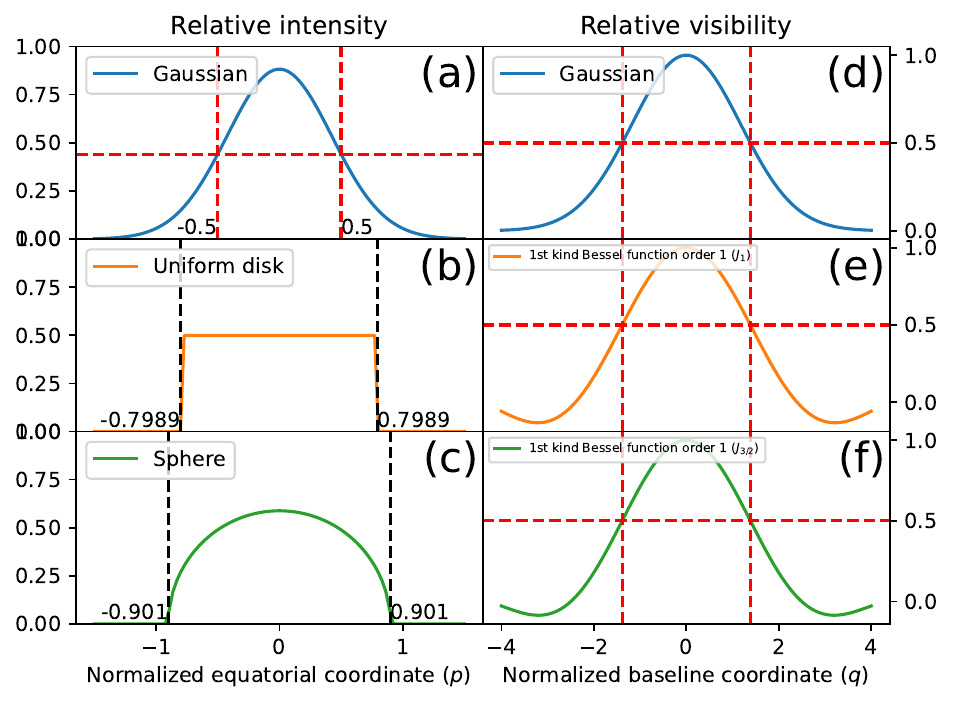}
    \caption{
    Theoretical visibility distributions (right panels from (a) to (c)) corresponding to intensity distributions (left panels from (d) to (f)).
    Red dashed lines denote FWHM and black dashed lines denote radius.
    Blue (panels~(a) and (d)), orange (panels~(b) and (e)), and green (panels~(c) and (f)) solid lines mean Gaussian, uniform disk, and sphere intensity distributions, respectively.
    All visibility distributions show the same FWHMs.
    }
    \label{fig:morphologies}
\end{figure}
\begin{table}[!ht]
    \caption{Visibilities of elliptic cylindrical symmetric morphologies.}
    \label{tab:morphologies_visibility}
    \centering
    \begin{tabular}{ll}
        \hline\hline
        Morphology & Visibility $V(q)$\\
        \hline
        Disk & $2\pi a\theta_{R}^{2}I_{0}\frac{J_{1}(q)}{q}$ \\
        Ellipsoid & $2\pi a\theta_{R}^{2}I_{0}\sqrt{\frac{\pi}{2}}\frac{J_{3/2}(q)}{q^{3/2}}$ \\
        Gaussian & $\frac{2\pi}{s^{2}}a\theta_{R}^{2}I_{0}\exp\Big(-\frac{q^{2}}{2s^{2}}\Big)$ \\
        \hline
    \end{tabular}
    \tablefoot
    {$J_{n}(x)$ is the $1^{\rm st}$ kind Bessel function of order, and $q$ is the normalized coordinates defined as Eq.~(\ref{eq:normalized_visibility_coordinate}).
    }
\end{table}
$q$ is the normalized coordinate defined as
\begin{equation}
    q\equiv2\pi\theta_{R}\sqrt{a^{2}u^{2}+v^{2}}=2\pi\theta_{R}\rho\sqrt{a^{2}\sin^{2}{\lambda}+\cos^{2}{\lambda}},
    \label{eq:normalized_visibility_coordinate}
\end{equation}
where $(u,v)$ is the baseline coordinate system, and $(\rho,\lambda)$ is the corresponding polar coordinate system.
It is assumed that the horizontal cross section bounded by the FWHM of the visibility functions will be the same between finite morphologies and a Gaussian morphology.
The ellipse equation of the finite morphology cross section bounded by the FWHM is
\begin{equation}
    q_{1/2}=2\pi\theta_{R}\rho\sqrt{a^{2}\sin^{2}{\lambda}+\cos^{2}{\lambda}},
    \label{eq:finite_morphology_cross_section}
\end{equation}
where $q_{1/2}=V^{-1}(V(0)/2)$, and $V^{-1}$ is the inverse function of $V$.
For the Gaussian morphology, $q_{1/2}=s\sqrt{2\ln{2}}$. When $\theta_{R}=\theta_{\rm FWHM}/2$, $q_{1/2}=2\ln{2}$.
Then, the ellipse equation of the Gaussian morphology cross section bounded by the FWHM is
\begin{equation}
    2\ln{2}=\pi\theta_{\rm FWHM}\rho\sqrt{a^{2}\sin^{2}{\lambda}+\cos^{2}{\lambda}}.
    \label{eq:Gaussian_morphology_cross_section}
\end{equation}
Under the assumption that the horizontal cross section bounded by the FWHM of the visibility functions is the same between finite morphologies and a Gaussian morphology, Eq.~(\ref{eq:finite_morphology_cross_section}) is equal to Eq.~(\ref{eq:Gaussian_morphology_cross_section}), yielding
\begin{equation}
    \theta_{R}=\frac{q_{1/2}}{4\ln{2}}\theta_{\rm FWHM}=M\theta_{\rm FWHM},
    \label{eq:size_scaling_factor}
\end{equation}
where the size scaling factor $M$ is defined as
\begin{equation}
    M\equiv\frac{q_{1/2}}{4\ln{2}}.
\end{equation}

\onecolumn
\section{Circular Gaussian fitting to 15~GHz VLBA data from MOJAVE}\label{sect:B}
In this section we describe the 2D circular Gaussian model fitting with the 15~GHz VLBA data.
The cell size of VLBI maps was set to $0.1\times0.1 ~{\rm mas}^{2}$ with $1024\times1024$ pixels, and a natural weighting is imposed.
We performed the model fitting with an increasing number of model components ($M$).
We stopped the fitting when the reduced $\chi^{2}$ improvement was less than 10~\%:
\begin{equation}
    \frac{1}{m}\frac{\chi_{{\rm vis}}^{2}(M-m)-\chi_{{\rm vis}}^{2}(M)}{\chi_{{\rm vis}}^{2}(M-m)}<0.1,
    \label{eq:chisquare_criteria}
\end{equation}
where $\chi_{\rm vis}^{2}(M)$ is the visibility reduced $\chi^{2}$ with respect to the number of model components $M$, and $m$ is the number of additional components.
The best fitting Gaussian model components were obtained only when the angular sizes FWHM ($\theta_{\rm FWHM}$) were larger than the minimum resolvable angular sizes even if the addition of another model component satisfied the reduced $\chi^{2}$ criteria.
The approximated 1$\sigma$ errors $\sigma_{\rm peak}$, $\sigma_{\rm tot}$, and $\sigma_{\rm FWHM}$ corresponding to $S_{\rm peak}$, $S_{\rm tot}$, and $\theta_{\rm FWHM}$, respectively, are determined by
\begin{equation}
    \sigma_{\rm peak}=\sigma_{\rm rms}\bigg(1+\frac{S_{\rm peak}}{\sigma_{\rm rms}}\bigg)^{1/2},\quad
    \sigma_{\rm tot}=\sigma_{\rm peak}\bigg(1+\frac{S_{\rm tot}^{2}}{S_{\rm peak}^2}\bigg)^{1/2},\quad
    \sigma_{\rm FWHM}=\theta_{\rm FWHM}\frac{\sigma_{\rm peak}}{S_{\rm peak}},
    \label{eq:Gaussian_model_error}
\end{equation}
where $S_{\rm peak}$, $S_{\rm tot}$, and $\sigma_{\rm rms}$ are the peak flux density, the total flux density, and the root-mean-square (RMS) of the model component, respectively.
We determined that the $\sigma_{\rm rms}$ calculation region is the combined area of the model component and the clean beam area constrained by the FWHM.
The minimum resolvable angular size $\theta_{\rm min}$ of a Gaussian component in a natural-weighting image is determined, following \citet{Lobanov2005}, as
\begin{equation}
    \theta_{\rm min}=\frac{4}{\pi}\bigg[\pi{\theta_{\rm FWHM}}^{2}\ln{2}\ln{\Big(\frac{\rm S/N}{{\rm S/N}-1}\Big)}\bigg]^{1/2},
    \label{eq:minimum_resolvable_size}
\end{equation}
where the signal-to-noise ratio (S/N) is defined by ${\rm S/N}=S_{\rm peak}/\sigma_{\rm rms}$.
\begin{longtable}{llllllll}
    \caption{Circular Gaussian model fitting parameters of 15~GHz VLBA data from MOJIVE.} \\
    \label{tab:VLBA_model_fitting_result} \\
    \hline\hline
    MJD & $M$ & $j$ & $S_{\rm tot}$~[Jy] & $\theta_{\rm FWHM}$~[mas] & $r$~[mas] & $\phi$~[deg] & $\chi_{\rm vis}^{2}$ \\
    \hline
    \endfirsthead
    \caption{Continued.} \\
    \hline\hline
    MJD & $M$ & $j$ & $S_{\rm tot}$~[Jy] & $\theta_{\rm FWHM}$~[mas] & $r$~[mas] & $\phi$~[deg] & $\chi_{\rm vis}^{2}$ \\
    \hline
    \endhead
    \hline
    \endfoot
    54838 & 3 & 0 & $0.346\pm0.012$ & $0.012\pm0.083$ & $0.083\pm0.002$ & $0.002\pm0.028$ & 1.080 \\ 
    ~ & ~ & 2 & $0.070\pm0.007$ & $0.007\pm1.145$ & $1.145\pm0.184$ & $0.184\pm0.152$ & ~ \\ 
    ~ & ~ & 3 & $0.107\pm0.005$ & $0.005\pm0.217$ & $0.217\pm0.011$ & $0.011\pm0.047$ & ~ \\ 
    54985 & 3 & 0 & $0.353\pm0.007$ & $0.007\pm0.095$ & $0.095\pm0.002$ & $0.002\pm0.016$ & 1.071 \\ 
    ~ & ~ & 2 & $0.080\pm0.008$ & $0.008\pm1.278$ & $1.278\pm0.247$ & $0.247\pm0.171$ & ~ \\ 
    ~ & ~ & 3 & $0.142\pm0.005$ & $0.005\pm0.309$ & $0.309\pm0.012$ & $0.012\pm0.034$ & ~ \\ 
    55389 & 2 & 0 & $0.279\pm0.030$ & $0.030\pm0.200$ & $0.200\pm0.020$ & $0.020\pm0.084$ & 1.249 \\ 
    ~ & ~ & 2 & $0.127\pm0.017$ & $0.017\pm1.105$ & $1.105\pm0.280$ & $0.280\pm0.211$ & ~ \\ 
    55513 & 2 & 0 & $0.250\pm0.028$ & $0.028\pm0.187$ & $0.187\pm0.020$ & $0.020\pm0.094$ & 1.244 \\ 
    ~ & ~ & 2 & $0.108\pm0.012$ & $0.012\pm1.086$ & $1.086\pm0.233$ & $0.233\pm0.190$ & ~ \\ 
    55619 & 2 & 0 & $0.234\pm0.034$ & $0.034\pm0.157$ & $0.157\pm0.021$ & $0.021\pm0.107$ & 1.186 \\ 
    ~ & ~ & 2 & $0.108\pm0.016$ & $0.016\pm1.289$ & $1.289\pm0.357$ & $0.357\pm0.218$ & ~ \\ 
    55963 & 2 & 0 & $0.241\pm0.024$ & $0.024\pm0.162$ & $0.162\pm0.015$ & $0.015\pm0.079$ & 1.202 \\ 
    ~ & ~ & 2 & $0.087\pm0.009$ & $0.009\pm1.165$ & $1.165\pm0.256$ & $0.256\pm0.188$ & ~ \\ 
    56351 & 2 & 0 & $0.246\pm0.025$ & $0.025\pm0.180$ & $0.180\pm0.017$ & $0.017\pm0.078$ & 1.154 \\ 
    ~ & ~ & 2 & $0.079\pm0.009$ & $0.009\pm1.472$ & $1.472\pm0.381$ & $0.381\pm0.212$ & ~ \\ 
    56682 & 3 & 0 & $0.286\pm0.013$ & $0.013\pm0.139$ & $0.139\pm0.006$ & $0.006\pm0.034$ & 1.194 \\ 
    ~ & ~ & 2 & $0.062\pm0.006$ & $0.006\pm1.585$ & $1.585\pm0.378$ & $0.378\pm0.196$ & ~ \\ 
    ~ & ~ & 3 & $0.064\pm0.003$ & $0.003\pm0.345$ & $0.345\pm0.021$ & $0.021\pm0.051$ & ~ \\ 
    57040 & 3 & 0 & $0.282\pm0.014$ & $0.014\pm0.158$ & $0.158\pm0.007$ & $0.007\pm0.039$ & 1.122 \\ 
    ~ & ~ & 3 & $0.064\pm0.004$ & $0.004\pm0.337$ & $0.337\pm0.026$ & $0.026\pm0.065$ & ~ \\ 
    ~ & ~ & 2 & $0.063\pm0.006$ & $0.006\pm1.297$ & $1.297\pm0.261$ & $0.261\pm0.168$ & ~ \\ 
    57271 & 2 & 0 & $0.283\pm0.028$ & $0.028\pm0.181$ & $0.181\pm0.016$ & $0.016\pm0.074$ & 1.429 \\ 
    ~ & ~ & 2 & $0.089\pm0.011$ & $0.011\pm1.282$ & $1.282\pm0.356$ & $0.356\pm0.222$ & ~ \\ 
    57409 & 2 & 0 & $0.221\pm0.020$ & $0.020\pm0.196$ & $0.196\pm0.017$ & $0.017\pm0.073$ & 1.413 \\ 
    ~ & ~ & 2 & $0.073\pm0.009$ & $0.009\pm1.123$ & $1.123\pm0.278$ & $0.278\pm0.209$ & ~ \\ 
    57555 & 3 & 0 & $0.232\pm0.006$ & $0.006\pm0.067$ & $0.067\pm0.001$ & $0.001\pm0.018$ & 1.222 \\ 
    ~ & ~ & 2 & $0.070\pm0.006$ & $0.006\pm1.332$ & $1.332\pm0.304$ & $0.304\pm0.176$ & ~ \\ 
    ~ & ~ & 3 & $0.123\pm0.004$ & $0.004\pm0.283$ & $0.283\pm0.011$ & $0.011\pm0.030$ & ~ \\ 
    57710 & 3 & 0 & $0.282\pm0.012$ & $0.012\pm0.048$ & $0.048\pm0.002$ & $0.002\pm0.031$ & 1.211 \\ 
    ~ & ~ & 2 & $0.062\pm0.007$ & $0.007\pm1.314$ & $1.314\pm0.334$ & $0.334\pm0.201$ & ~ \\ 
    ~ & ~ & 3 & $0.173\pm0.006$ & $0.006\pm0.312$ & $0.312\pm0.011$ & $0.011\pm0.029$ & ~ \\ 
    57921 & 2 & 0 & $0.429\pm0.026$ & $0.026\pm0.106$ & $0.106\pm0.005$ & $0.005\pm0.040$ & 1.636 \\ 
    ~ & ~ & 3 & $0.117\pm0.009$ & $0.009\pm0.487$ & $0.487\pm0.050$ & $0.050\pm0.080$ & ~ \\ 
    58230 & 3 & 0 & $0.636\pm0.017$ & $0.017\pm0.096$ & $0.096\pm0.002$ & $0.002\pm0.017$ & 1.387 \\ 
    ~ & ~ & 3 & $0.154\pm0.007$ & $0.007\pm0.262$ & $0.262\pm0.012$ & $0.012\pm0.038$ & ~ \\ 
    ~ & ~ & 2 & $0.094\pm0.010$ & $0.010\pm1.321$ & $1.321\pm0.319$ & $0.319\pm0.191$ & ~ \\ 
    58269 & 3 & 0 & $0.736\pm0.025$ & $0.025\pm0.097$ & $0.097\pm0.003$ & $0.003\pm0.020$ & 1.384 \\ 
    ~ & ~ & 3 & $0.139\pm0.009$ & $0.009\pm0.273$ & $0.273\pm0.019$ & $0.019\pm0.055$ & ~ \\ 
    ~ & ~ & 2 & $0.099\pm0.010$ & $0.010\pm1.421$ & $1.421\pm0.341$ & $0.341\pm0.186$ & ~ \\ 
    58468 & 3 & 0 & $0.788\pm0.028$ & $0.028\pm0.113$ & $0.113\pm0.003$ & $0.003\pm0.022$ & 1.337 \\ 
    ~ & ~ & 3 & $0.139\pm0.008$ & $0.008\pm0.309$ & $0.309\pm0.019$ & $0.019\pm0.051$ & ~ \\ 
    ~ & ~ & 2 & $0.080\pm0.007$ & $0.007\pm1.355$ & $1.355\pm0.254$ & $0.254\pm0.155$ & ~ \\ 
    58699 & 2 & 0 & $1.606\pm0.144$ & $0.144\pm0.120$ & $0.120\pm0.007$ & $0.007\pm0.043$ & 6.312 \\ 
    ~ & ~ & 3 & $0.119\pm0.019$ & $0.019\pm0.334$ & $0.334\pm0.060$ & $0.060\pm0.135$ & ~ \\ 
    58834 & 2 & 0 & $1.906\pm0.157$ & $0.157\pm0.138$ & $0.138\pm0.007$ & $0.007\pm0.041$ & 9.546 \\ 
    ~ & ~ & 3 & $0.162\pm0.022$ & $0.022\pm0.443$ & $0.443\pm0.071$ & $0.071\pm0.133$ & ~ \\ 
    58895 & 3 & 0 & $1.699\pm0.134$ & $0.134\pm0.131$ & $0.131\pm0.006$ & $0.006\pm0.042$ & 6.091 \\ 
    ~ & ~ & 3 & $0.406\pm0.038$ & $0.038\pm0.309$ & $0.309\pm0.027$ & $0.027\pm0.077$ & ~ \\ 
    ~ & ~ & 2 & $0.056\pm0.009$ & $0.009\pm0.244$ & $0.244\pm0.040$ & $0.040\pm0.142$ & ~ \\ 
    58948 & 2 & 0 & $1.901\pm0.088$ & $0.088\pm0.131$ & $0.131\pm0.004$ & $0.004\pm0.030$ & 5.734 \\ 
    ~ & ~ & 3 & $0.377\pm0.030$ & $0.030\pm0.510$ & $0.510\pm0.041$ & $0.041\pm0.089$ & ~ \\ 
    58977 & 3 & 0 & $1.704\pm0.141$ & $0.141\pm0.142$ & $0.142\pm0.007$ & $0.007\pm0.044$ & 7.995 \\ 
    ~ & ~ & 3 & $0.503\pm0.046$ & $0.046\pm0.304$ & $0.304\pm0.025$ & $0.025\pm0.070$ & ~ \\ 
    ~ & ~ & 2 & $0.061\pm0.012$ & $0.012\pm0.207$ & $0.207\pm0.042$ & $0.042\pm0.170$ & ~ \\ 
    59013 & 2 & 0 & $1.638\pm0.165$ & $0.165\pm0.182$ & $0.182\pm0.012$ & $0.012\pm0.048$ & 7.187 \\ 
    ~ & ~ & 3 & $0.476\pm0.046$ & $0.046\pm0.493$ & $0.493\pm0.055$ & $0.055\pm0.082$ & ~ \\ 
    59062 & 3 & 0 & $1.390\pm0.138$ & $0.138\pm0.173$ & $0.173\pm0.012$ & $0.012\pm0.051$ & 5.865 \\ 
    ~ & ~ & 3 & $0.597\pm0.049$ & $0.049\pm0.461$ & $0.461\pm0.039$ & $0.039\pm0.065$ & ~ \\ 
    ~ & ~ & 2 & $0.086\pm0.016$ & $0.016\pm1.156$ & $1.156\pm0.378$ & $0.378\pm0.247$ & ~ \\ 
    59207 & 3 & 0 & $1.208\pm0.130$ & $0.130\pm0.169$ & $0.169\pm0.013$ & $0.013\pm0.064$ & 8.117 \\ 
    ~ & ~ & 3 & $0.633\pm0.067$ & $0.067\pm0.387$ & $0.387\pm0.038$ & $0.038\pm0.083$ & ~ \\ 
    ~ & ~ & 2 & $0.181\pm0.033$ & $0.033\pm1.314$ & $1.314\pm0.457$ & $0.457\pm0.292$ & ~ \\ 
    59275 & 3 & 0 & $0.921\pm0.089$ & $0.089\pm0.228$ & $0.228\pm0.017$ & $0.017\pm0.059$ & 4.320 \\ 
    ~ & ~ & 3 & $0.478\pm0.040$ & $0.040\pm0.436$ & $0.436\pm0.038$ & $0.038\pm0.068$ & ~ \\ 
    ~ & ~ & 2 & $0.166\pm0.024$ & $0.024\pm1.372$ & $1.372\pm0.455$ & $0.455\pm0.255$ & ~ \\ 
    59433 & 3 & 0 & $0.529\pm0.043$ & $0.043\pm0.134$ & $0.134\pm0.009$ & $0.009\pm0.056$ & 2.989 \\ 
    ~ & ~ & 2 & $0.220\pm0.025$ & $0.025\pm1.289$ & $1.289\pm0.281$ & $0.281\pm0.177$ & ~ \\ 
    ~ & ~ & 3 & $0.324\pm0.028$ & $0.028\pm0.355$ & $0.355\pm0.031$ & $0.031\pm0.073$ & ~ \\ 
    59634 & 3 & 0 & $0.629\pm0.046$ & $0.046\pm0.122$ & $0.122\pm0.007$ & $0.007\pm0.043$ & 3.090 \\ 
    ~ & ~ & 2 & $0.230\pm0.026$ & $0.026\pm1.195$ & $1.195\pm0.302$ & $0.302\pm0.179$ & ~ \\ 
    ~ & ~ & 3 & $0.234\pm0.024$ & $0.024\pm0.312$ & $0.312\pm0.034$ & $0.034\pm0.078$ & ~ \\ 
    59741 & 4 & 0 & $0.411\pm0.018$ & $0.018\pm0.070$ & $0.070\pm0.003$ & $0.003\pm0.031$ & 1.667 \\ 
    ~ & ~ & 2 & $0.181\pm0.015$ & $0.015\pm1.246$ & $1.246\pm0.214$ & $0.214\pm0.137$ & ~ \\ 
    ~ & ~ & 3 & $0.237\pm0.011$ & $0.011\pm0.285$ & $0.285\pm0.013$ & $0.013\pm0.036$ & ~ \\ 
    ~ & ~ & 1 & $0.027\pm0.002$ & $0.002\pm2.313$ & $2.313\pm0.744$ & $0.744\pm0.253$ & ~ \\ 
    59979 & 4 & 0 & $0.648\pm0.028$ & $0.028\pm0.076$ & $0.076\pm0.003$ & $0.003\pm0.031$ & 1.385 \\ 
    ~ & ~ & 2 & $0.095\pm0.007$ & $0.007\pm1.095$ & $1.095\pm0.146$ & $0.146\pm0.123$ & ~ \\ 
    ~ & ~ & 3 & $0.102\pm0.005$ & $0.005\pm0.274$ & $0.274\pm0.015$ & $0.015\pm0.049$ & ~ \\ 
    ~ & ~ & 1 & $0.034\pm0.002$ & $0.002\pm2.932$ & $2.932\pm0.840$ & $0.840\pm0.261$ & ~ \\ 
    60126 & 4 & 0 & $0.811\pm0.055$ & $0.055\pm0.154$ & $0.154\pm0.008$ & $0.008\pm0.048$ & 3.279 \\ 
    ~ & ~ & 3 & $0.315\pm0.018$ & $0.018\pm0.319$ & $0.319\pm0.018$ & $0.018\pm0.053$ & ~ \\ 
    ~ & ~ & 2 & $0.086\pm0.011$ & $0.011\pm0.889$ & $0.889\pm0.176$ & $0.176\pm0.182$ & ~ \\ 
    ~ & ~ & 1 & $0.054\pm0.005$ & $0.005\pm3.656$ & $3.656\pm1.318$ & $1.318\pm0.328$ & ~ \\ 
    60262 & 3 & 0 & $0.572\pm0.028$ & $0.028\pm0.076$ & $0.076\pm0.003$ & $0.003\pm0.033$ & 2.292 \\ 
    ~ & ~ & 3 & $0.432\pm0.019$ & $0.019\pm0.327$ & $0.327\pm0.014$ & $0.014\pm0.036$ & ~ \\ 
    ~ & ~ & 2 & $0.095\pm0.009$ & $0.009\pm1.156$ & $1.156\pm0.247$ & $0.247\pm0.176$ & ~ \\ 
\end{longtable}
\tablefoot{
$S_{\rm tot}$ is the total flux density, $\theta_{\rm FWHM}$ is the FWHM of a model component in image domain, $r$ is the position radius, $\phi$ is the position angle, and $\chi_{\rm vis}^{2}$ is the visibility reduced $\chi^{2}$. $M$ is the number of model components and $j$ is a model index, where $j=0$, 1, 2, and 3 are matched to VLBA~C0, J1, J2, and J3 components.
}

\section{Flare decomposition}\label{sect:C}
The total number of decomposed flares is determined by trans-dimensional Bayesian inference~\citep{Brewer2014}.
Let $S[k]$ be the observed flux density of light curves and $\sigma_{S}[k]$ the 1$\sigma$ error at $k^{{\rm th}}$ epoch.
We assume that a light curve consists of exponential flares as follows~\citep{Kang2021}
\begin{equation}
    F_{i}(t)=2F_{0,i}\Bigg[\exp{\Big(\frac{t_{0,i}-t}{\tau_{i}}\Big)}+\exp{\Big(\frac{t-t_{0,i}}{s_{i}\tau_{i}}\Big)}\Bigg]^{-1}.
    \label{eq:flare_model_function}
\end{equation}
Each flare has four parameters, the reference time $t_{0,i}$, the reference flux density $F_{0,i}=F_{i}(t_{0,i})$, the rising timescale $\tau_{i}$, and the skewness $s_{i}$, which is the ratio of decaying to rising timescale for the $i^{{\rm th}}$ flare.
The multi-flare model is constructed as
\begin{equation}
    F(t)=\sum_{i=1}^{N}F_{i}(t)+F_{{\rm qs}},
    \label{eq:total_flare_model_function}
\end{equation}
where $N$ is the number of flares and $F_{{\rm qs}}$ is the quiescent flux density.
The peak time {$t_{{\rm p},i}$} and the peak flux density {$F_{{\rm p},i}$} of each flare are derived by the local maxima of Eq.~(\ref{eq:flare_model_function}) as
\begin{equation}
    t_{{\rm p},i}=t_{0,i}+\frac{s_{i}\tau}{s_{i}+1}\ln s_{i},
    \label{eq:flare_peak_time}
\end{equation}
\begin{equation}
    F_{{\rm p},i}=2F_{0,i}\Big[s_{i}^{-s_{i}/(s_{i}+1)}+s_{i}^{1/(s_{i}+1)}\Big]^{-1}.
    \label{eq:flare_peak_flux_density}
\end{equation}
We introduced hyper parameters that describe prior distributions
of the flare parameters~\citep{Brewer2014}.
$\mu_{F_{0}}$, $\mu_{\tau}$, and $\mu_{s}$ are means of prior distributions for $F_{0,i}$, $\tau_{i}$, and $s_{i}$, and $\sigma_{s}$ is the half width of a uniform prior distribution for $s_{i}$.
The likelihood $\mathcal{L}$ is derived from the joint independent normal distribution as
\begin{equation}
    \mathcal{L}=(2\pi)^{-\frac{K}{2}}\prod_{k=1}^{K}\frac{1}{\sigma_{S}[k]}\exp{\Bigg[-\frac{1}{2}\bigg(\frac{S[k]-F(t[k])}{\sigma_{S}[k]}\bigg)^{2}\Bigg]}
    \label{eq:likelihood_function},
\end{equation}
where $t[k]$ is a date of $k^{\rm th}$ epoch, and $K$ is the total number of epochs.
We chose the maximum likelihood sample in posterior samples as a point estimation.
{We estimated the Gaussian kernel density of posterior samples using the Python library \texttt{scikit-learn}~\citep{Pedregosa2011} to estimate the 1$\sigma$ Bayesian credible interval considered as the 1$\sigma$ error of parameters.}
When posterior samples include several numbers of components $N$, we selected the minimum reduced $\chi^{2}$ result.
We calculate the 1$\sigma$ error of $t_{{\rm p},i}$ and $F_{{\rm p},i}$ by using the Python package \texttt{uncertainties}.
\begin{table}[!h]
    \caption{Flare decomposition parameters.}
    \label{tab:flare_decomposition_parameters}
    \centering
    \begin{tabular}{lllllll@{\extracolsep{10pt}}l@{\extracolsep{5pt}}l}
        \hline\hline
        Data & Flare name & \multicolumn{5}{c}{Model parameters\tablefootmark{a}} & \multicolumn{2}{c}{Peak parameters\tablefootmark{b}} \\
        \cline{3-7}\cline{8-9}
        ~ & ~ & $F_{\rm qs}$~[Jy] & $t_{0}$~[MJD] & $\tau$~[day]& $F_{0}$~[Jy] & $s$ & $t_{\rm p}$~[MJD] & $F_{\rm p}$~[Jy] \\ 
        \hline
        VLBA~C0 & C0a & $0.164_{-0.086}^{+0.086}$ & $54779_{-133}^{+81}$ & $163_{-90}^{+61}$ & $0.17_{-0.07}^{+0.09}$ & $3.1_{-0.7}^{+1.1}$ & $54917_{-170}^{+134}$ & $0.20_{-0.09}^{+0.11}$ \\ 
        ~ & C0b & ~ & $56896_{-101}^{+1114}$ & $386_{-232}^{+96}$ & $0.14_{-0.08}^{+0.19}$ & $0.8_{-0.4}^{+0.8}$ & $56849_{-123}^{+1121}$ & $0.14_{-0.09}^{+0.19}$ \\ 
        ~ & C0c & ~ & $58079_{-40}^{+137}$ & $178_{-34}^{+162}$ & $0.40_{-0.17}^{+0.18}$ & $3.1_{-0.9}^{+0.9}$ & $58230_{-100}^{+215}$ & $0.46_{-0.21}^{+0.22}$ \\ 
        ~ & C0d & ~ & $58758_{-20}^{+33}$ & $125_{-62}^{+17}$ & $1.44_{-0.21}^{+0.20}$ & $2.7_{-0.6}^{+1.4}$ & $58850_{-73}^{+99}$ & $1.61_{-0.41}^{+0.50}$ \\ 
        ~ & C0e & ~ & $60059_{-27}^{+53}$ & $92_{-23}^{+18}$ & $0.54_{-0.11}^{+0.12}$ & $2.0_{-0.9}^{+0.9}$ & $60101_{-55}^{+72}$ & $0.58_{-0.19}^{+0.19}$ \\ 
        \hline
        VLBA~J2 & J2a & $0.064_{-0.005}^{+0.004}$ & $55374_{-69}^{+233}$ & $239_{-79}^{+199}$ & $0.05_{-0.02}^{+0.02}$ & $1.6_{-0.9}^{+0.7}$ & $55440_{-140}^{+258}$ & $0.05_{-0.02}^{+0.02}$ \\ 
        ~ & J2b & ~ & $58382_{-150}^{+1033}$ & $142_{-44}^{+177}$ & $0.05_{-0.03}^{+0.05}$ & $0.4_{-0.3}^{+1.0}$ & $58345_{-153}^{+1034}$ & $0.05_{-0.04}^{+0.05}$ \\ 
        ~ & J2c & ~ & $59445_{-55}^{+68}$ & $133_{-20}^{+28}$ & $0.18_{-0.04}^{+0.04}$ & $1.9_{-0.6}^{+0.6}$ & $59501_{-74}^{+83}$ & $0.19_{-0.06}^{+0.06}$ \\ 
        \hline
        VLBA~J3 & J3a & $0.000_{-0.000}^{+0.000}$ & $55303_{-263}^{+91}$ & $347_{-59}^{+117}$ & $0.22_{-0.08}^{+0.09}$ & $1.8_{-0.3}^{+0.9}$ & $55440_{-273}^{+209}$ & $0.23_{-0.10}^{+0.11}$ \\ 
        ~ & J3b & ~ & $57620_{-2}^{+76}$ & $326_{-75}^{+90}$ & $0.13_{-0.01}^{+0.01}$ & $3.8_{-0.9}^{+0.5}$ & $57963_{-173}^{+163}$ & $0.15_{-0.02}^{+0.02}$ \\ 
        ~ & J3c & ~ & $59026_{-16}^{+10}$ & $98_{-12}^{+6}$ & $0.46_{-0.04}^{+0.04}$ & $3.9_{-0.4}^{+0.5}$ & $59132_{-30}^{+28}$ & $0.55_{-0.06}^{+0.07}$ \\ 
        ~ & J3d & ~ & $60160_{-7}^{+9}$ & $30_{-1}^{+3}$ & $0.44_{-0.08}^{+0.10}$ & $4.3_{-1.0}^{+0.1}$ & $60196_{-16}^{+12}$ & $0.54_{-0.13}^{+0.13}$ \\ 
        \hline
        OVRO~SD & SD1 & $0.281_{-0.001}^{+0.001}$ & $53563.8_{-0.7}^{+1.7}$ & $511.6_{-0.7}^{+6.9}$ & $0.182_{-0.113}^{+0.114}$ & $2.42_{-0.17}^{+0.19}$ & $53883.0_{-51.6}^{+56.6}$ & $0.199_{-0.123}^{+0.125}$ \\ 
        + & SD2 & ~ & $54554.7_{-0.2}^{+0.1}$ & $2.9_{-0.2}^{+514.4}$ & $0.244_{-0.082}^{+0.062}$ & $0.80_{-0.02}^{+0.09}$ & $54554.4_{-0.2}^{+50.8}$ & $0.246_{-0.082}^{+0.062}$ \\ 
        VLBA~TC & SD3 & ~ & $54701.0_{-6.5}^{+9.9}$ & $14.8_{-0.9}^{+62.9}$ & $0.040_{-0.018}^{+0.074}$ & $2.51_{-1.72}^{+0.47}$ & $54710.7_{-15.4}^{+43.0}$ & $0.044_{-0.025}^{+0.082}$ \\ 
        ~ & SD4~(C0a) & ~ & $54716.3_{-6.8}^{+1.7}$ & $145.1_{-142.4}^{+4.7}$ & $0.176_{-0.076}^{+0.077}$ & $2.51_{-0.25}^{+0.17}$ & $54811.6_{-96.0}^{+15.7}$ & $0.193_{-0.084}^{+0.085}$ \\ 
        ~ & SD5 & ~ & $55143.2_{-0.2}^{+1.6}$ & $72.0_{-3.1}^{+4.7}$ & $0.155_{-0.048}^{+0.021}$ & $1.19_{-0.13}^{+0.14}$ & $55150.1_{-5.0}^{+5.8}$ & $0.155_{-0.049}^{+0.022}$ \\ 
        ~ & SD6 & ~ & $55547.3_{-0.2}^{+221.5}$ & $8.4_{-2.3}^{+0.4}$ & $0.140_{-0.051}^{+0.052}$ & $2.11_{-0.32}^{+1.82}$ & $55551.6_{-3.2}^{+221.6}$ & $0.149_{-0.070}^{+0.085}$ \\ 
        ~ & SD7 & ~ & $55773.6_{-637.0}^{+5.1}$ & $73.3_{-1.6}^{+1.3}$ & $0.052_{-0.027}^{+0.043}$ & $3.63_{-0.39}^{+0.18}$ & $55847.8_{-637.2}^{+10.3}$ & $0.061_{-0.032}^{+0.051}$ \\ 
        ~ & SD8 & ~ & $56559.9_{-0.2}^{+585.0}$ & $78.6_{-0.3}^{+2.2}$ & $0.151_{-0.055}^{+0.042}$ & $2.49_{-0.19}^{+0.19}$ & $56611.0_{-8.7}^{+585.0}$ & $0.166_{-0.061}^{+0.047}$ \\ 
        ~ & SD9 & ~ & $57137.5_{-1146.4}^{+0.5}$ & $60.3_{-51.8}^{+1.2}$ & $0.198_{-0.054}^{+0.008}$ & $1.73_{-0.19}^{+0.19}$ & $57158.6_{-1146.5}^{+6.6}$ & $0.206_{-0.058}^{+0.015}$ \\ 
        ~ & SD10 & ~ & $57682.8_{-0.3}^{+5.8}$ & $52.4_{-47.8}^{+2.5}$ & $0.095_{-0.060}^{+0.060}$ & $1.16_{-0.13}^{+0.15}$ & $57686.9_{-5.3}^{+7.2}$ & $0.095_{-0.060}^{+0.060}$ \\ 
        ~ & SD11 & ~ & $57756.1_{-611.9}^{+4.0}$ & $152.7_{-0.4}^{+0.4}$ & $0.210_{-0.056}^{+0.003}$ & $3.28_{-0.18}^{+0.19}$ & $57895.3_{-612.0}^{+16.7}$ & $0.244_{-0.065}^{+0.010}$ \\ 
        ~ & SD12 & ~ & $57994.1_{-316.7}^{+1.2}$ & $27.1_{-2.6}^{+131.3}$ & $0.067_{-0.013}^{+0.049}$ & $1.03_{-0.23}^{+0.17}$ & $57994.5_{-316.7}^{+3.1}$ & $0.067_{-0.014}^{+0.050}$ \\ 
        ~ & SD13 & ~ & $58228.0_{-0.7}^{+4.6}$ & $27.2_{-5.0}^{+0.8}$ & $0.106_{-0.051}^{+0.051}$ & $3.34_{-0.20}^{+0.21}$ & $58253.3_{-5.6}^{+5.6}$ & $0.124_{-0.059}^{+0.060}$ \\ 
        ~ & SD14~(C0c) & ~ & $58307.3_{-0.2}^{+1.6}$ & $154.2_{-2.3}^{+2.6}$ & $0.345_{-0.076}^{+0.007}$ & $1.57_{-0.22}^{+0.18}$ & $58349.5_{-19.4}^{+16.2}$ & $0.354_{-0.082}^{+0.024}$ \\ 
        ~ & SD15 & ~ & $58407.7_{-0.3}^{+0.3}$ & $4.0_{-0.3}^{+0.3}$ & $0.123_{-0.072}^{+0.088}$ & $0.97_{-0.14}^{+0.20}$ & $58407.6_{-0.4}^{+0.5}$ & $0.123_{-0.072}^{+0.088}$ \\ 
        ~ & SD16 & ~ & $58468.2_{-0.2}^{+0.4}$ & $3.2_{-0.2}^{+1.6}$ & $0.088_{-0.054}^{+0.088}$ & $2.31_{-0.82}^{+0.37}$ & $58470.1_{-1.5}^{+1.3}$ & $0.095_{-0.061}^{+0.096}$ \\ 
        ~ & SD17 & ~ & $58538.8_{-1.7}^{+252.2}$ & $37.1_{-0.3}^{+90.9}$ & $0.332_{-0.155}^{+0.159}$ & $3.43_{-0.20}^{+0.18}$ & $58574.2_{-4.4}^{+266.8}$ & $0.389_{-0.182}^{+0.187}$ \\ 
        ~ & SD18 & ~ & $58706.0_{-0.3}^{+0.2}$ & $1.1_{-0.2}^{+0.2}$ & $0.190_{-0.062}^{+0.109}$ & $1.58_{-0.30}^{+1.79}$ & $58706.3_{-0.4}^{+1.1}$ & $0.195_{-0.093}^{+0.145}$ \\ 
        ~ & SD19~(C0d) & ~ & $58790.0_{-0.2}^{+0.4}$ & $128.0_{-0.3}^{+0.2}$ & $1.469_{-0.085}^{+0.004}$ & $3.69_{-0.09}^{+0.01}$ & $58921.7_{-5.5}^{+2.6}$ & $1.750_{-0.104}^{+0.015}$ \\ 
        ~ & SD20 & ~ & $59003.1_{-0.2}^{+197.9}$ & $59.5_{-0.9}^{+2.4}$ & $0.210_{-0.086}^{+0.087}$ & $0.98_{-0.13}^{+0.13}$ & $59002.5_{-3.8}^{+197.9}$ & $0.210_{-0.087}^{+0.087}$ \\ 
        ~ & SD21 & ~ & $59163.7_{-0.2}^{+469.7}$ & $35.7_{-22.5}^{+0.8}$ & $0.106_{-0.024}^{+0.077}$ & $1.57_{-0.16}^{+1.95}$ & $59173.6_{-10.3}^{+471.4}$ & $0.109_{-0.048}^{+0.096}$ \\ 
        ~ & SD22 & ~ & $59203.0_{-0.1}^{+873.1}$ & $13.2_{-0.2}^{+0.2}$ & $0.349_{-0.048}^{+0.048}$ & $3.84_{-0.11}^{+0.08}$ & $59217.1_{-0.8}^{+873.1}$ & $0.420_{-0.058}^{+0.058}$ \\ 
        ~ & SD23 & ~ & $59632.6_{-1.4}^{+541.3}$ & $25.5_{-0.2}^{+0.2}$ & $0.279_{-0.012}^{+0.047}$ & $2.77_{-0.19}^{+0.19}$ & $59651.7_{-3.1}^{+541.3}$ & $0.313_{-0.019}^{+0.055}$ \\ 
        ~ & SD24~(C0e) & ~ & $60078.2_{-0.2}^{+1.1}$ & $77.0_{-26.5}^{+1.2}$ & $0.418_{-0.078}^{+0.004}$ & $3.81_{-0.12}^{+0.09}$ & $60159.8_{-28.5}^{+4.2}$ & $0.501_{-0.094}^{+0.009}$ \\ 
        ~ & SD25 & ~ & $60170.7_{-2.9}^{+0.5}$ & $141.6_{-1.3}^{+1.3}$ & $0.308_{-0.146}^{+0.149}$ & $1.05_{-0.15}^{+0.17}$ & $60174.4_{-11.0}^{+12.7}$ & $0.308_{-0.147}^{+0.150}$ \\ 
        \hline
    \end{tabular}
    \tablefoot{
    The flares are listed in ascending order of $t_{0}$.
    $F_{\rm qs}$ is the quiescent flux density, $t_{0}$ is the reference time, $\tau$ is the rising timescale, $F_{0}$ is the reference flux density at $t_{0}$, and $s_{i}$ is the skewness which is the ratio of decaying to rising timescale.
    $t_{\rm p}$ and $F_{\rm p}$ are the peak time and flux density, respectively.
    \tablefoottext{a}{The component model parameters ($t_{0}$, $\tau$, $F_{0}$, and $s$) and quiescent flux density ($F_{\rm qs}$) are given from Eqs.~(\ref{eq:flare_model_function}) and (\ref{eq:total_flare_model_function}), respectively.}
    \tablefoottext{b}{The peak time ($t_{\rm p}$) and peak flux density ($F_{\rm p}$) are given from Eqs.~(\ref{eq:flare_peak_time}) and (\ref{eq:flare_peak_flux_density}), respectively.}
    Flares cross-identified between the OVRO~SD+VLBA~TC and individual VLBA components are indicated in parentheses next to the flare name.
    }
\end{table}

\onecolumn
\section{Distance measurements using all VLBA and OVRO~SD+VLBA~TC flares in all VLBA epochs}\label{sect:D}
We summarized the distance estimates using all VLBA and OVRO~SD+VLBA~TC flares in all VLBA epochs in Tables~\ref{tab:distances_C0+SD} and \ref{tab:distances_J2+J3}.
\begin{table}[!h]
    \centering
    \caption{Distance measurements using flares of VLBA~C0 and OVRO~SD+VLBA~TC light curves.}
    \label{tab:distances_C0+SD}
    \begin{tabular}{lp{1.48cm}p{1.48cm}p{1.48cm}p{1.48cm}p{1.48cm}p{1.48cm}p{1.48cm}p{1.48cm}p{1.48cm}}
        \hline\hline
        MJD & \multicolumn{9}{c}{Distance~[Mpc]} \\
        \cline{2-10}
        ~ & C0a & C0b & C0c & C0d & C0e & SD4 & SD14 & SD19 & SD24 \\ 
        ~ & ~ & ~ & ~ & ~ & ~ & (C0a) & (C0c) & (C0d) & (C0e) \\
        ~ & (1) & (2) & (3) & (4) & (5) & (6) & (7) & (8) & (9) \\
        \hline
        54838 & $748_{-540}^{+517}$ & $1287_{-1106}^{+1798}$ & $1902_{-981}^{+1971}$ & $4713_{-2663}^{+1664}$ & $1245_{-525}^{+498}$ & $660_{-711}^{+296}$ & $1283_{-318}^{+145}$ & $5273_{-567}^{+474}$ & $908_{-364}^{+84}$ \\ 
        54985 & $487_{-350}^{+335}$ & $839_{-718}^{+1170}$ & $1240_{-633}^{+1282}$ & $3072_{-1722}^{+1062}$ & $811_{-337}^{+319}$ & $430_{-462}^{+191}$ & $836_{-198}^{+73}$ & $3437_{-275}^{+187}$ & $592_{-234}^{+35}$ \\ 
        55389 & $53_{-41}^{+39}$ & $91_{-82}^{+129}$ & $134_{-79}^{+144}$ & $333_{-211}^{+151}$ & $88_{-45}^{+43}$ & $47_{-52}^{+25}$ & $91_{-34}^{+28}$ & $372_{-114}^{+112}$ & $64_{-32}^{+19}$ \\ 
        55513 & $65_{-51}^{+49}$ & $111_{-101}^{+159}$ & $164_{-98}^{+177}$ & $407_{-261}^{+190}$ & $107_{-56}^{+54}$ & $57_{-64}^{+31}$ & $111_{-43}^{+36}$ & $455_{-147}^{+144}$ & $78_{-39}^{+25}$ \\ 
        55619 & $110_{-90}^{+87}$ & $189_{-179}^{+275}$ & $280_{-181}^{+310}$ & $693_{-477}^{+366}$ & $183_{-105}^{+103}$ & $97_{-111}^{+58}$ & $189_{-88}^{+77}$ & $775_{-316}^{+313}$ & $133_{-75}^{+54}$ \\ 
        55963 & $99_{-76}^{+73}$ & $171_{-153}^{+243}$ & $253_{-146}^{+270}$ & $626_{-389}^{+274}$ & $165_{-82}^{+79}$ & $88_{-97}^{+45}$ & $170_{-61}^{+48}$ & $700_{-197}^{+192}$ & $121_{-58}^{+33}$ \\ 
        56351 & $73_{-56}^{+54}$ & $125_{-113}^{+178}$ & $185_{-107}^{+198}$ & $459_{-286}^{+203}$ & $121_{-60}^{+58}$ & $64_{-71}^{+34}$ & $125_{-45}^{+36}$ & $513_{-147}^{+144}$ & $88_{-43}^{+25}$ \\ 
        56682 & $158_{-115}^{+110}$ & $272_{-235}^{+381}$ & $403_{-210}^{+419}$ & $998_{-570}^{+362}$ & $263_{-113}^{+108}$ & $140_{-151}^{+64}$ & $272_{-71}^{+38}$ & $1116_{-152}^{+137}$ & $192_{-79}^{+24}$ \\ 
        57040 & $108_{-79}^{+75}$ & $185_{-161}^{+260}$ & $274_{-144}^{+285}$ & $679_{-391}^{+251}$ & $179_{-78}^{+74}$ & $95_{-103}^{+44}$ & $185_{-50}^{+29}$ & $760_{-116}^{+107}$ & $131_{-54}^{+19}$ \\ 
        57271 & $72_{-55}^{+53}$ & $124_{-111}^{+176}$ & $183_{-105}^{+195}$ & $452_{-281}^{+198}$ & $119_{-59}^{+57}$ & $63_{-70}^{+33}$ & $123_{-44}^{+35}$ & $506_{-141}^{+138}$ & $87_{-42}^{+24}$ \\ 
        57409 & $56_{-43}^{+41}$ & $96_{-86}^{+137}$ & $143_{-81}^{+152}$ & $353_{-217}^{+151}$ & $93_{-45}^{+43}$ & $49_{-55}^{+25}$ & $96_{-33}^{+25}$ & $395_{-104}^{+101}$ & $68_{-32}^{+17}$ \\ 
        57555 & $1419_{-1020}^{+977}$ & $2442_{-2094}^{+3409}$ & $3610_{-1850}^{+3735}$ & $8943_{-5026}^{+3113}$ & $2362_{-986}^{+934}$ & $1253_{-1347}^{+558}$ & $2435_{-586}^{+234}$ & $10007_{-899}^{+680}$ & $1722_{-684}^{+123}$ \\ 
        57710 & $3900_{-2829}^{+2712}$ & $6711_{-5791}^{+9390}$ & $9921_{-5173}^{+10308}$ & $24577_{-14013}^{+8876}$ & $6491_{-2779}^{+2641}$ & $3443_{-3716}^{+1568}$ & $6691_{-1734}^{+909}$ & $27501_{-3619}^{+3236}$ & $4733_{-1934}^{+567}$ \\ 
        57921 & $359_{-263}^{+252}$ & $617_{-536}^{+866}$ & $913_{-484}^{+953}$ & $2261_{-1308}^{+846}$ & $597_{-262}^{+250}$ & $317_{-343}^{+148}$ & $616_{-171}^{+103}$ & $2530_{-415}^{+387}$ & $435_{-183}^{+67}$ \\ 
        58230 & $478_{-344}^{+329}$ & $823_{-705}^{+1148}$ & $1216_{-623}^{+1258}$ & $3013_{-1692}^{+1047}$ & $796_{-332}^{+314}$ & $422_{-454}^{+188}$ & $820_{-197}^{+77}$ & $3371_{-293}^{+216}$ & $580_{-230}^{+39}$ \\ 
        58269 & $466_{-336}^{+322}$ & $803_{-689}^{+1121}$ & $1186_{-610}^{+1228}$ & $2939_{-1656}^{+1030}$ & $776_{-325}^{+308}$ & $412_{-443}^{+184}$ & $800_{-195}^{+83}$ & $3289_{-322}^{+258}$ & $566_{-226}^{+46}$ \\ 
        58468 & $290_{-209}^{+200}$ & $500_{-429}^{+698}$ & $738_{-380}^{+765}$ & $1829_{-1031}^{+642}$ & $483_{-203}^{+192}$ & $256_{-276}^{+115}$ & $498_{-122}^{+53}$ & $2047_{-204}^{+165}$ & $352_{-141}^{+29}$ \\ 
        58699 & $246_{-181}^{+174}$ & $424_{-369}^{+595}$ & $626_{-336}^{+655}$ & $1551_{-905}^{+592}$ & $410_{-183}^{+174}$ & $217_{-236}^{+103}$ & $422_{-121}^{+78}$ & $1736_{-313}^{+296}$ & $299_{-128}^{+51}$ \\ 
        58834 & $162_{-118}^{+113}$ & $278_{-242}^{+390}$ & $412_{-218}^{+429}$ & $1020_{-589}^{+380}$ & $269_{-118}^{+112}$ & $143_{-155}^{+66}$ & $278_{-76}^{+45}$ & $1141_{-182}^{+170}$ & $196_{-82}^{+30}$ \\ 
        58895 & $187_{-137}^{+131}$ & $322_{-279}^{+452}$ & $476_{-252}^{+496}$ & $1179_{-680}^{+438}$ & $312_{-136}^{+130}$ & $165_{-179}^{+77}$ & $321_{-88}^{+52}$ & $1320_{-208}^{+193}$ & $227_{-95}^{+34}$ \\ 
        58948 & $187_{-135}^{+129}$ & $322_{-276}^{+449}$ & $476_{-245}^{+493}$ & $1178_{-665}^{+414}$ & $311_{-131}^{+124}$ & $165_{-178}^{+74}$ & $321_{-79}^{+34}$ & $1319_{-134}^{+109}$ & $227_{-91}^{+19}$ \\ 
        58977 & $147_{-107}^{+103}$ & $252_{-219}^{+354}$ & $373_{-198}^{+389}$ & $923_{-535}^{+346}$ & $244_{-107}^{+102}$ & $129_{-140}^{+60}$ & $251_{-70}^{+42}$ & $1033_{-171}^{+160}$ & $178_{-75}^{+28}$ \\ 
        59013 & $71_{-52}^{+50}$ & $121_{-106}^{+171}$ & $179_{-98}^{+189}$ & $445_{-263}^{+175}$ & $117_{-54}^{+51}$ & $62_{-68}^{+30}$ & $121_{-37}^{+25}$ & $498_{-102}^{+97}$ & $86_{-37}^{+17}$ \\ 
        59062 & $81_{-61}^{+58}$ & $140_{-123}^{+197}$ & $207_{-113}^{+218}$ & $513_{-304}^{+203}$ & $136_{-62}^{+59}$ & $72_{-78}^{+35}$ & $140_{-43}^{+30}$ & $574_{-120}^{+115}$ & $99_{-43}^{+20}$ \\ 
        59207 & $88_{-66}^{+64}$ & $152_{-134}^{+215}$ & $225_{-125}^{+237}$ & $557_{-335}^{+228}$ & $147_{-69}^{+66}$ & $78_{-85}^{+39}$ & $152_{-49}^{+36}$ & $623_{-145}^{+140}$ & $107_{-48}^{+24}$ \\ 
        59275 & $35_{-27}^{+26}$ & $61_{-54}^{+86}$ & $90_{-50}^{+95}$ & $224_{-135}^{+91}$ & $59_{-28}^{+27}$ & $31_{-34}^{+15}$ & $61_{-20}^{+14}$ & $250_{-58}^{+56}$ & $43_{-19}^{+10}$ \\ 
        59433 & $174_{-130}^{+124}$ & $300_{-263}^{+422}$ & $443_{-242}^{+466}$ & $1097_{-652}^{+436}$ & $290_{-133}^{+127}$ & $154_{-168}^{+75}$ & $299_{-92}^{+64}$ & $1228_{-261}^{+251}$ & $211_{-93}^{+43}$ \\ 
        59634 & $231_{-171}^{+164}$ & $398_{-348}^{+559}$ & $588_{-317}^{+616}$ & $1458_{-854}^{+561}$ & $385_{-173}^{+165}$ & $204_{-222}^{+97}$ & $397_{-116}^{+76}$ & $1631_{-306}^{+290}$ & $281_{-121}^{+50}$ \\ 
        59741 & $1236_{-896}^{+859}$ & $2127_{-1834}^{+2975}$ & $3144_{-1637}^{+3265}$ & $7788_{-4436}^{+2805}$ & $2057_{-879}^{+835}$ & $1091_{-1177}^{+496}$ & $2120_{-547}^{+283}$ & $8714_{-1124}^{+1000}$ & $1500_{-612}^{+175}$ \\ 
        59979 & $978_{-707}^{+677}$ & $1683_{-1448}^{+2352}$ & $2488_{-1288}^{+2580}$ & $6163_{-3493}^{+2193}$ & $1628_{-690}^{+655}$ & $863_{-930}^{+390}$ & $1678_{-423}^{+204}$ & $6896_{-803}^{+693}$ & $1187_{-479}^{+122}$ \\ 
        60126 & $117_{-86}^{+82}$ & $201_{-175}^{+282}$ & $297_{-158}^{+310}$ & $736_{-426}^{+276}$ & $194_{-86}^{+82}$ & $103_{-112}^{+48}$ & $200_{-56}^{+34}$ & $824_{-137}^{+128}$ & $142_{-60}^{+22}$ \\ 
        60262 & $976_{-708}^{+679}$ & $1679_{-1449}^{+2350}$ & $2482_{-1295}^{+2580}$ & $6149_{-3508}^{+2223}$ & $1624_{-696}^{+661}$ & $862_{-930}^{+393}$ & $1674_{-435}^{+229}$ & $6881_{-914}^{+819}$ & $1184_{-484}^{+143}$ \\ 
        \hline
    \end{tabular}
    \tablefoot{
    In columns (1) to (5) and columns (6) to (9) are names of decomposed flares using the VLBA~C0 and VLBA~C0+OVRO+TC light curves, respectively.
    For each columns, distances are calculated using a timescale and a peak flux densities of each flare in columns (1) to (9).
    For each row, distances are calculated using an angular size of the VLBA~C0 which is measured in each epoch noted in the 1st column by MJD.
    }
\end{table}
\begin{table}[!h]
    \centering
    \caption{Distance measurements using flares of VLBA~J2 and J3 light curves.}
    \label{tab:distances_J2+J3}
    \begin{tabular}{llllllll}
        \hline\hline
        MJD & \multicolumn{7}{c}{Distance~[Mpc]} \\
        \cline{2-8}
        ~ & J2a & J2b & J2c & J3a & J3b & J3c & J3d \\
        ~ & (1) & (2) & (3) & (4) & (5) & (6) & (7) \\ 
        \hline
        54838 & $0.11_{-0.08}^{+0.11}$ & $0.06_{-0.06}^{+0.11}$ & $0.23_{-0.13}^{+0.14}$ & $103.74_{-49.58}^{+61.28}$ & $64.48_{-20.60}^{+22.73}$ & $70.93_{-15.64}^{+14.30}$ & $21.20_{-5.87}^{+6.48}$ \\ 
        54985 & $0.08_{-0.06}^{+0.09}$ & $0.05_{-0.04}^{+0.08}$ & $0.16_{-0.11}^{+0.11}$ & $35.90_{-16.80}^{+20.91}$ & $22.31_{-6.79}^{+7.56}$ & $24.54_{-4.85}^{+4.33}$ & $7.34_{-1.90}^{+2.13}$ \\ 
        55389 & $0.12_{-0.11}^{+0.15}$ & $0.07_{-0.08}^{+0.13}$ & $0.25_{-0.21}^{+0.21}$ & - & - & - & - \\ 
        55513 & $0.13_{-0.11}^{+0.15}$ & $0.08_{-0.08}^{+0.13}$ & $0.27_{-0.19}^{+0.20}$ & - & - & - & - \\ 
        55619 & $0.08_{-0.08}^{+0.10}$ & $0.05_{-0.05}^{+0.08}$ & $0.16_{-0.14}^{+0.14}$ & - & - & - & - \\ 
        55963 & $0.10_{-0.09}^{+0.12}$ & $0.06_{-0.06}^{+0.11}$ & $0.22_{-0.16}^{+0.16}$ & - & - & - & - \\ 
        56351 & $0.05_{-0.05}^{+0.06}$ & $0.03_{-0.03}^{+0.05}$ & $0.11_{-0.09}^{+0.09}$ & - & - & - & - \\ 
        56682 & $0.04_{-0.04}^{+0.05}$ & $0.02_{-0.03}^{+0.04}$ & $0.09_{-0.07}^{+0.07}$ & $25.68_{-12.56}^{+15.40}$ & $15.96_{-5.37}^{+5.87}$ & $17.55_{-4.28}^{+3.99}$ & $5.25_{-1.55}^{+1.70}$ \\ 
        57040 & $0.08_{-0.06}^{+0.08}$ & $0.04_{-0.04}^{+0.08}$ & $0.16_{-0.11}^{+0.11}$ & $27.51_{-14.00}^{+16.95}$ & $17.10_{-6.23}^{+6.73}$ & $18.81_{-5.29}^{+5.02}$ & $5.62_{-1.84}^{+1.98}$ \\ 
        57271 & $0.08_{-0.08}^{+0.10}$ & $0.05_{-0.05}^{+0.08}$ & $0.16_{-0.15}^{+0.15}$ & - & - & - & - \\ 
        57409 & $0.12_{-0.11}^{+0.14}$ & $0.07_{-0.07}^{+0.12}$ & $0.24_{-0.20}^{+0.20}$ & - & - & - & - \\ 
        57555 & $0.07_{-0.06}^{+0.08}$ & $0.04_{-0.04}^{+0.07}$ & $0.14_{-0.11}^{+0.11}$ & $46.64_{-21.83}^{+27.18}$ & $28.99_{-8.83}^{+9.83}$ & $31.89_{-6.32}^{+5.64}$ & $9.53_{-2.47}^{+2.76}$ \\ 
        57710 & $0.07_{-0.07}^{+0.09}$ & $0.04_{-0.05}^{+0.08}$ & $0.15_{-0.13}^{+0.13}$ & $34.83_{-16.26}^{+20.27}$ & $21.65_{-6.56}^{+7.31}$ & $23.81_{-4.66}^{+4.14}$ & $7.12_{-1.83}^{+2.05}$ \\ 
        57921 & - & - & - & $9.17_{-5.04}^{+5.96}$ & $5.70_{-2.39}^{+2.53}$ & $6.27_{-2.19}^{+2.11}$ & $1.87_{-0.73}^{+0.76}$ \\ 
        58230 & $0.07_{-0.06}^{+0.08}$ & $0.04_{-0.04}^{+0.07}$ & $0.15_{-0.12}^{+0.12}$ & $58.61_{-27.88}^{+34.51}$ & $36.43_{-11.51}^{+12.73}$ & $40.07_{-8.63}^{+7.85}$ & $11.98_{-3.27}^{+3.62}$ \\ 
        58269 & $0.06_{-0.05}^{+0.07}$ & $0.03_{-0.04}^{+0.06}$ & $0.12_{-0.09}^{+0.10}$ & $52.17_{-26.05}^{+31.73}$ & $32.43_{-11.38}^{+12.36}$ & $35.67_{-9.41}^{+8.85}$ & $10.66_{-3.33}^{+3.61}$ \\ 
        58468 & $0.07_{-0.05}^{+0.07}$ & $0.04_{-0.04}^{+0.07}$ & $0.14_{-0.09}^{+0.09}$ & $35.65_{-17.44}^{+21.39}$ & $22.16_{-7.45}^{+8.15}$ & $24.38_{-5.95}^{+5.54}$ & $7.29_{-2.16}^{+2.35}$ \\ 
        58699 & - & - & - & $28.47_{-20.10}^{+22.40}$ & $17.70_{-10.79}^{+11.11}$ & $19.47_{-10.98}^{+10.85}$ & $5.82_{-3.43}^{+3.51}$ \\ 
        58834 & - & - & - & $12.14_{-8.01}^{+9.06}$ & $7.55_{-4.20}^{+4.35}$ & $8.30_{-4.20}^{+4.13}$ & $2.48_{-1.32}^{+1.36}$ \\ 
        58895 & $11.31_{-8.21}^{+11.88}$ & $6.64_{-6.07}^{+11.18}$ & $23.45_{-13.92}^{+14.34}$ & $35.90_{-18.85}^{+22.59}$ & $22.31_{-8.62}^{+9.24}$ & $24.54_{-7.60}^{+7.27}$ & $7.34_{-2.58}^{+2.75}$ \\ 
        58948 & - & - & - & $7.98_{-4.11}^{+4.96}$ & $4.96_{-1.85}^{+1.99}$ & $5.46_{-1.60}^{+1.52}$ & $1.63_{-0.55}^{+0.59}$ \\ 
        58977 & $18.62_{-15.00}^{+20.61}$ & $10.93_{-10.69}^{+18.79}$ & $38.60_{-26.60}^{+27.19}$ & $37.71_{-19.50}^{+23.49}$ & $23.44_{-8.80}^{+9.47}$ & $25.78_{-7.63}^{+7.28}$ & $7.70_{-2.62}^{+2.80}$ \\ 
        59013 & - & - & - & $8.79_{-4.97}^{+5.83}$ & $5.46_{-2.40}^{+2.54}$ & $6.01_{-2.25}^{+2.18}$ & $1.80_{-0.74}^{+0.77}$ \\ 
        59062 & $0.11_{-0.12}^{+0.14}$ & $0.06_{-0.08}^{+0.12}$ & $0.22_{-0.23}^{+0.23}$ & $10.78_{-5.61}^{+6.74}$ & $6.70_{-2.55}^{+2.73}$ & $7.37_{-2.22}^{+2.12}$ & $2.20_{-0.76}^{+0.81}$ \\ 
        59207 & $0.07_{-0.09}^{+0.10}$ & $0.04_{-0.06}^{+0.08}$ & $0.15_{-0.16}^{+0.17}$ & $18.29_{-9.87}^{+11.73}$ & $11.37_{-4.61}^{+4.91}$ & $12.50_{-4.17}^{+4.02}$ & $3.74_{-1.40}^{+1.48}$ \\ 
        59275 & $0.06_{-0.07}^{+0.09}$ & $0.04_{-0.05}^{+0.07}$ & $0.13_{-0.14}^{+0.14}$ & $12.76_{-6.69}^{+8.02}$ & $7.93_{-3.06}^{+3.28}$ & $8.73_{-2.69}^{+2.57}$ & $2.61_{-0.91}^{+0.97}$ \\ 
        59433 & $0.08_{-0.06}^{+0.09}$ & $0.05_{-0.05}^{+0.08}$ & $0.16_{-0.12}^{+0.12}$ & $23.68_{-12.46}^{+14.93}$ & $14.72_{-5.71}^{+6.11}$ & $16.19_{-5.04}^{+4.83}$ & $4.84_{-1.71}^{+1.82}$ \\ 
        59634 & $0.10_{-0.09}^{+0.12}$ & $0.06_{-0.06}^{+0.10}$ & $0.20_{-0.17}^{+0.17}$ & $34.64_{-19.36}^{+22.80}$ & $21.53_{-9.29}^{+9.83}$ & $23.68_{-8.63}^{+8.36}$ & $7.08_{-2.84}^{+2.98}$ \\ 
        59741 & $0.09_{-0.06}^{+0.09}$ & $0.05_{-0.05}^{+0.08}$ & $0.18_{-0.11}^{+0.11}$ & $45.74_{-21.67}^{+26.87}$ & $28.43_{-8.91}^{+9.86}$ & $31.27_{-6.61}^{+5.99}$ & $9.35_{-2.52}^{+2.80}$ \\ 
        59979 & $0.13_{-0.08}^{+0.13}$ & $0.07_{-0.06}^{+0.12}$ & $0.26_{-0.14}^{+0.14}$ & $51.22_{-24.65}^{+30.40}$ & $31.83_{-10.34}^{+11.37}$ & $35.02_{-7.98}^{+7.34}$ & $10.47_{-2.96}^{+3.26}$ \\ 
        60126 & $0.23_{-0.19}^{+0.26}$ & $0.14_{-0.13}^{+0.24}$ & $0.49_{-0.33}^{+0.34}$ & $32.46_{-15.76}^{+19.37}$ & $20.17_{-6.67}^{+7.32}$ & $22.19_{-5.25}^{+4.86}$ & $6.63_{-1.92}^{+2.10}$ \\ 
        60262 & $0.11_{-0.09}^{+0.12}$ & $0.06_{-0.06}^{+0.11}$ & $0.22_{-0.16}^{+0.16}$ & $30.12_{-14.22}^{+17.65}$ & $18.72_{-5.82}^{+6.45}$ & $20.60_{-4.27}^{+3.85}$ & $6.16_{-1.64}^{+1.82}$ \\
        \hline
    \end{tabular}
    \tablefoot{
    In columns (1) to (4) and columns (5) to (7) are names of decomposed flares using the VLBA~J2 and VLBA~J3 light curves, respectively.
    For each column, distances are calculated using a timescale and a peak flux densities of each flare in columns (1) to (7).
    For each row, distances in columns (1) to (4) and columns (5) to (7) are calculated using angular sizes of the VLBA~J2 and VLBA~J3, respectively, which are measured in each epoch noted in the 1st column by MJD.
    }
\end{table}

\onecolumn
\section{Parameter uncertainties and offsets}\label{sect:E}
\begin{figure}[!h]
    \centering
    \includegraphics[width=1\linewidth]{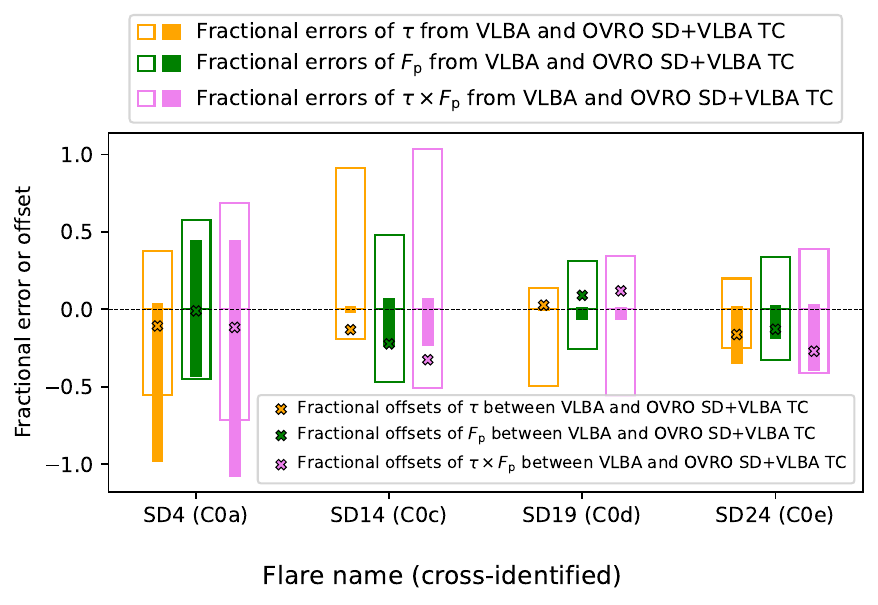}
    \caption{Fractional errors and offsets of decomposition parameters (timescales in orange, peak flux densities in green, and their multiplications in pink).
    Fractional errors are plotted as open boxes for VLBA and filled boxes for OVRO~SD+VLBA~TC, and fractional offsets are cross symbols.
    }
    \label{fig:fractional_errors}
\end{figure}
To investigate the effect of the systematic error from the OVRO~SD+VLBA~TC on estimating the model parameters of the flare decomposition (Table~\ref{tab:flare_decomposition_parameters}) and the distances, we compared the parameter offsets (the timescale and peak flux density) between VLBA and OVRO~SD+VLBA~TC with the corresponding parameter uncertainties of VLBA~C0 flares (C0a, C0c, C0d, and C0e) as shown in Fig.~\ref{fig:fractional_errors}.
The fractional offsets (cross symbols) of the parameters ($\tau$, $F_{\rm p}$, and $\tau\times F_{\rm p}$) overlap with the fractional errors (open boxes) from VLBA.
We found that the offsets in the flares are within the corresponding parameter uncertainties of VLBA~C0.

\section{The difference between standard e-folding timescales and decomposition timescales}\label{sect:F}
In order to compare the e-folding timescale of a selected period in individual decomposed flare (e.g., VLBA~C0) with the corresponding timescales of decomposed flare, we investigated the derivatives ($\partial\ln{F(t)}/\partial t$) of a simulated flare Eq.~(\ref{eq:flare_model_function}).
\begin{equation}
    \frac{\partial\ln{F(t)}}{\partial t}=-\frac{1}{\tau}\frac{-\exp{(\frac{t_{0}-t}{\tau})}+\frac{1}{s}\exp{(\frac{t_{0}-t}{s\tau})}}{\exp{(\frac{t_{0}-t}{\tau})}+\exp{(\frac{t_{0}-t}{s\tau})}}
    \label{eq:flare_model_function_derivative}
\end{equation}
As a result, we found that the derivative (Eq.~(\ref{eq:flare_model_function_derivative})) changes as a function of time with approaching $1/\tau~(t\ll t_{0})$ and $-1/s\tau~(t\gg t_{0})$ and converging to zero at the peak of the flare, implying that the e-folding timescale (similar to the inverse derivative ($\partial\ln{F(t)}/\partial t)^{-1}$) becomes larger than the decomposed timescale ($\tau$) in particular at the near the peak of the flare.
Therefore, flares should always be decomposed.

\end{appendix}

\end{document}